\documentclass[%
 reprint,
superscriptaddress,
nofootinbib,
 amsmath,amssymb,
 aps,
 prd,
]{revtex4-2}

\usepackage{amssymb, amsmath}
\usepackage[utf8]{inputenc}
\usepackage{subfigure}
\usepackage{comment}
\usepackage{graphicx}

\usepackage{enumerate}
\usepackage{booktabs}
\usepackage{lipsum}
\usepackage{xspace}
\usepackage{aas_macros}
\usepackage{multirow}
\usepackage{lineno}
\usepackage[dvipsnames]{xcolor}
\usepackage{url}
\usepackage{hyperref}
\hypersetup{
    colorlinks = true,
    citecolor = {MidnightBlue},
    linkcolor = {BrickRed},
    urlcolor = {BrickRed}
}

\newcommand{\nv}{\hat{\bf n}}
\newcommand{\wtj}[6]{\left(\begin{array}{ccc} #1 & #2 & #3\\#4 & #5 & #6\end{array} \right)}
\newcommand{\nmt}{{\tt NaMaster}\xspace}

\begin{document}

\title{Analytical covariances for catalogue-based pseudo-\texorpdfstring{$C_\ell$}{CL}s}

\author{Kevin Wolz}
\email{kevin.wolz@physics.ox.ac.uk}
\affiliation{Department of Physics, University of Oxford, Denys Wilkinson Building, Keble Road, Oxford OX1 3RH, United Kingdom}
\author{Elyas Farah}
\affiliation{Argelander-Institut für Astronomie, Universität Bonn, Auf dem Hügel 71, D-53121 Bonn, Germany}
\author{Robert Reischke}
\affiliation{Argelander-Institut für Astronomie, Universität Bonn, Auf dem Hügel 71, D-53121 Bonn, Germany}
\author{David Alonso}
\affiliation{Department of Physics, University of Oxford, Denys Wilkinson Building, Keble Road, Oxford OX1 3RH, United Kingdom}
\author{Andrina Nicola}
\affiliation{Jodrell Bank Centre for Astrophysics, Department of Physics and Astronomy, The University of Manchester, Manchester M13 9PL, UK}

\begin{abstract}
  Multiple cosmological observables, such as the galaxy overdensity or cosmic shear, consist of fields sampled at the discrete positions of astrophysical sources. Recent work has presented methods to estimate the angular power spectra of such fields, avoiding the construction of pixelated sky maps and the finite-resolution effects associated with them. In this work, we present a method to estimate the disconnected (also known as ``Gaussian'') covariance of these angular power spectra, addressing subtle effects such as the effective area overlap between different catalogue-based fields and the additional Poisson-like variance arising from the discrete nature of the catalogues. The method relies on the so-called Narrow-Kernel Approximation to account for the contribution of distinct source pairs to the estimator, while including the noise-like contributions from self-pairs exactly. We explicitly compare this approach with a brute-force method that can produce the exact covariance for sparse samples, and validate it against simulations. We show that the method is accurate in realistic scenarios, spanning both dense and noise-dominated datasets (e.g., cosmic shear) and sparse, noise-dominated observables (e.g., fast radio bursts). The method is implemented in the public code \nmt.
\end{abstract}

\maketitle

\section{Introduction}\label{sec:intro}
    Projected observations of cosmological fields contain a wealth of information about the anisotropies in the late-time Universe, the growth of structure, and the distribution of baryons in the interstellar and intergalactic medium. Some of the most relevant examples of such observables in astrophysics are the primary anisotropies of the Cosmic Microwave Background (CMB) \cite{1807.06209,2503.14452}, galaxy clustering \cite{1912.08209,2309.06443}, gravitational lensing of the CMB \cite{astro-ph/0601594,2304.05202} and galaxy shapes \cite{astro-ph/9912508,2501.07938}, the kinematic and thermal Sunyaev-Zel'dovich effects \cite{astro-ph/9808050,1811.02310,2407.07152,2512.14625}, galaxy velocities \cite{2410.06229,2504.02525,2605.15947}, or the dispersion measure of fast radio bursts \cite{2506.08932,2509.05866,2602.12174,2604.22105}. This wealth of probes has also stimulated the study of cross-correlations and joint analyses that combine different observables to break cosmological and astrophysical degeneracies, and avoid additive systematic biases \cite{2105.12108,2309.03258,2407.04607,2503.24385,2603.04269,2606.28099}.
    
    The efficient analysis of projected fields on the sky relies on robust estimators with known statistical properties, such as their mean and covariance. An example is the pseudo-$C_\ell$ (PCL) estimator of the angular power spectrum, a fast, powerful, and therefore widely used tool to analyse auto- and cross-correlations on the partial sky \cite{astro-ph/0105302,1809.09603}. In addition to measuring the power spectra themselves, obtaining accurate estimates of their covariance matrix is a key part of any cosmological analysis, ensuring that all statistical uncertainties can be robustly propagated to the final parameter constraints. In this context, the use of analytical estimators for the covariance matrix has become ubiquitous in cosmological analyses due to the increasing number of power spectra involved in joint-probe studies and the correspondingly large number of high-fidelity simulations needed to obtain accurate sample covariances. Analytical PCL covariances have been studied within the CMB and large-scale-structure (LSS) communities, where accurate estimators have been proposed for the disconnected trispectrum of scalar and polarised Gaussian fields \citep{astro-ph/0012087,astro-ph/0410394,1906.11765,2010.09717,2204.13721} and, in the LSS case, the contributions from the connected non-Gaussian trispectrum and super-sample mode-coupling \citep{1711.07467}. The dominant contribution for most cosmological probes is typically the disconnected ``Gaussian'' part, with the non-Gaussian terms contributing at the $\sim5$-10\% level \citep{1807.04266}. Developing accurate estimates of the disconnected Gaussian covariance for power spectra is therefore of paramount importance.

    Analytical approaches to power spectrum covariances on reasonably small scales ($\ell\gtrsim1000$) invariably involve numerical approximations to simplify what would otherwise be an $O(\ell_{\rm max}^6)$ calculation. Different approaches have been proposed in the literature \cite{2012.08568,2204.13721}, including hybrid estimators that incorporate simulations to improve estimator accuracy in the presence of survey-specific noise properties \cite{2412.07068}. A popular approach is the so-called Narrow Kernel Approximation (NKA), first introduced in \cite{astro-ph/0307515,astro-ph/0410394} for CMB data, and more recently improved and systematised for general projected fields in \cite{1906.11765,2010.09717}. The NKA has become a widely used tool in many cosmological analyses \cite{2304.00701,2309.05659,2410.22397,2412.07068,2510.18981} due to its accuracy and high computational efficiency, which make it particularly well-suited for small-scale analyses ($\ell\gtrsim10^3$).
    
    Traditionally, power spectrum-based studies have relied on the harmonic-space analysis of sky maps, in which a given field is sampled on a regular pixel grid. This is the natural approach to analysing map-based observables, such as the intensity and polarisation of the sky across different frequency bands (e.g., in CMB or intensity-mapping experiments). However, multiple cosmological probes of the LSS consist of measurements of astrophysical quantities at the discrete positions of a catalogue of sources. Examples of such catalogue-based fields include cosmic shear, the dispersion measure of fast radio bursts (FRBs), galaxy clustering, and the galaxy momentum density field \cite{2512.14625,2605.15947}. The traditional approach to studying these probes is to bin the catalogue into pixels, which leads to a number of numerical effects, including aliasing, the introduction of an effective pixel window function that is difficult to characterise, and the unstable statistical coupling between different angular scales due to the complexity of the associated sky geometry \citep{2025JCAP...05..048H}. Recently, estimators for the angular power spectra of such catalogue-based fields have been introduced that avoid constructing pixel-based maps \cite{2312.12285,2407.21013,2408.16903} by exploiting recent developments in irregularly sampled spherical harmonic transforms (SHTs) \cite{2304.10431}. The resulting estimator is free from finite-resolution effects, bias, and numerical instability arising from catalogue shot noise.
        
    In this work, we generalise the improved NKA (iNKA) approach of \cite{2010.09717} by developing an estimator for the power-spectrum covariance of catalogue-based fields. Our estimator separates the contributions from distinct source pairs and self-pairs. As we will show, the former can be approximated using the standard iNKA approach, treating each source as a finite-sized ``cloud'' whose size is determined by the catalogue density and the angular scales we wish to recover. The contribution from self-pairs can, in turn, be calculated exactly as an effective source of white noise. The resulting estimator retains the numerical complexity of the standard iNKA and the PCL estimator itself, $O(\ell_{\rm max}^3)$, while remaining highly accurate and robust over a wide range of catalogue densities and field sensitivities. We run a validation suite of 32 representative cross- and auto-field combinations of catalogue- and map-based fields, demonstrating consistent and excellent agreement with empirical covariances from simulations. We also present a highly parallelised implementation of the exact ``brute-force'' method based on the direct sum of all source pairs, applicable to low-density catalogues on relatively large scales.

    This paper is structured as follows: Section \ref{sec:pcls} reviews the main results associated with the PCL estimator, including the standard iNKA approach and the catalogue-based estimator. Section \ref{sec:theo} presents the extension of the iNKA approach to catalogue-based fields and the direct-sum calculation. Our method is then validated in different regimes in Section \ref{sec:val}. We conclude in Section \ref{sec:conc}. 

\section{Crash course on pseudo-\texorpdfstring{$C_\ell$}{CL}s}\label{sec:pcls}
  We begin with a brief introduction to the pseudo-$C_\ell$ estimator to set the scene, establish the notation used in the rest of the paper, and clarify the challenges encountered in extending the estimation of power-spectrum covariances to the case of discretely sampled fields. For simplicity, we focus our discussion in the main text on scalar fields, and we summarise the modifications to the method presented here for spin-$s$ fields in Appendix \ref{app:spin_s}. We refer readers to previous papers presenting this formalism for further details. In particular, \cite{1809.09603} presents the general pseudo-$C_\ell$ estimator, \cite{1906.11765} and \cite{2010.09717} present the iNKA used to estimate pseudo-$C_\ell$ covariances, and \cite{2407.21013} introduces catalogue-based power spectra.

  \subsection{Pseudo-\texorpdfstring{$C_\ell$}{CL}s}\label{ssec:pcls.pcl}
    Consider a field $a(\nv)$ defined on the sphere, where the unit vector $\nv$ denotes a given position on the celestial sphere. In all practical cases, we observe a ``masked'' version of this field $\tilde{a}\equiv\,w_a\,a$, where $w_a(\nv)$ is the sky mask associated with $a$. In the simplest case, $w_a$ is a binary map that simply defines the regions of the sky where the field has been observed. More generally, $w_a$ can be thought of as a weights map that optimally combines the different sky regions to maximise the signal-to-noise ratio of the field's power spectrum. The SHT of $\tilde{a}$ and $a$ are related to each other through a convolution:
    \begin{equation}\label{eq:mask_harmonic}
      \tilde{a}_{\ell m}=\sum_{\ell'm'}{\cal M}_{\ell m,\ell'm'}(w_a)\,a_{\ell'm'},
    \end{equation}
    where the map-level coupling coefficients ${\cal M}$, depending on the sky mask, are
    \begin{equation}\label{eq:map_mcm}
      {\cal M}_{\ell m,\ell'm'}(w_a)\equiv\int d\nv\,w_a(\nv)\,Y^*_{\ell m}(\nv)\,Y_{\ell'm'}(\nv).
    \end{equation}

    In the case of full-sky observations ($w_a=1$ throughout, and ${\cal M}_{\ell m,\ell'm'}(w_a)=\delta^K_{\ell\ell'}\delta^K_{mm'}$, where $\delta^K_{ij}$ is the Kronecker delta), the optimal estimator for the power spectrum of two fields, $a$ and $b$, is
    \begin{equation}
      \hat{C}^{ab}_\ell=\frac{1}{2\ell+1}\sum_m{\rm Re}(a^*_{\ell m}b_{\ell m}).
    \end{equation}
    The pseudo-$C_\ell$ estimator consists of na\"ively applying this to the masked fields instead:
    \begin{equation}\label{eq:pcl_def}
      \tilde{C}^{ab}_\ell\equiv\frac{1}{2\ell+1}\sum_m{\rm Re}(\tilde{a}^*_{\ell m}\tilde{b}_{\ell m}).
    \end{equation}
    However, just as the presence of a mask induces a coupling between different angular multipoles at the map level (Eq. \ref{eq:mask_harmonic}), it also induces mode coupling at the level of the power spectrum. Indeed, the ensemble average of the pseudo-$C_\ell$ estimator is:
    \begin{equation}\label{eq:pcl}
      \left\langle \tilde{C}^{ab}_\ell\right\rangle=\sum_{\ell'} M_{\ell\ell'}(w_a,w_b)\,C^{ab}_{\ell'},
    \end{equation}
    where $C^{ab}_\ell$ is the true underlying power spectrum, and $M_{\ell\ell'}(w_a,w_b)$ is the \emph{mode-coupling matrix}. The latter depends only on the pseudo-$C_\ell$ of the two masks, $\tilde{C}^{w_a,w_b}_\ell$, and can be calculated analytically with modest computational complexity \citep{astro-ph/0105302} ($\sim{\cal O}(\ell_{\rm max}^3)$). Specifically, for spin-0 quantities, $M_{\ell\ell'}(w_a,w_b)=(2\ell'+1)\Xi_{\ell\ell'}(w_a,w_b)$, where the coupling coefficients $\Xi_{\ell\ell'}$ are
    \begin{equation}\label{eq:Xi}
      \Xi_{\ell\ell'}(w_a,w_b)\equiv\sum_{\ell''}\frac{2\ell''+1}{4\pi}\,\tilde{C}^{w_a,w_b}_{\ell''}\,\wtj{\ell}{\ell'}{\ell''}{0}{0}{0}^2,
    \end{equation}
    involving the Wigner-$3j$ symbols.

  \subsection{Gaussian covariances}\label{ssec:pcls.cov}
    The statistical uncertainties of the estimated power spectra are described by their covariance matrix ${\rm Cov}(\tilde{C}^{ab}_\ell,\tilde{C}^{cd}_{\ell'})$. A dominant component of the power spectrum covariance for many cosmological fields, including most large-scale structure tracers, is the so-called ``disconnected'' or ``Gaussian'' covariance \cite{1607.00043}, corresponding to the disconnected component of the fields' trispectrum (the only non-zero component for Gaussian random fields). The Gaussian pseudo-$C_\ell$ covariance is thus:
    \begin{equation}\nonumber
      {\rm Cov}(\tilde{C}^{ab}_\ell,\tilde{C}^{cd}_{\ell'})=\sum_{mm'}\frac{\langle \tilde{a}_{\ell m}\tilde{c}_{\ell'm'}^*\rangle\langle \tilde{b}^*_{\ell m}\tilde{d}_{\ell'm'}\rangle+(c\leftrightarrow d)}{(2\ell+1)(2\ell'+1)},
    \end{equation}
    where $(c\leftrightarrow d)$ denotes exchanging the roles of fields $c$ and $d$. 
    
    The key to estimating the Gaussian covariance is then the calculation of general two-point correlators of the form $\langle\tilde{a}_{\ell m}\tilde{c}_{\ell'm'}^*\rangle$. Using Eq. \ref{eq:mask_harmonic}, we can write this term as:
    \begin{equation}\label{eq:general_2pt}
      \langle\tilde{a}_{\ell m}\tilde{c}_{\ell'm'}^*\rangle=
      \sum_{\ell''m''}{\cal M}_{\ell m,\ell''m''}(w_a)\,{\cal M}^{*}_{\ell' m',\ell''m''}(w_c)C_{\ell''}^{ac}.
    \end{equation}
    The standard Narrow Kernel Approximation (NKA) now proceeds by exploiting that, for masks with a reasonable angular extent, ${\cal M}_{\ell m,\ell''m''}(w)$ is narrowly peaked around $(\ell,m)=(\ell'',m'')$, and thus the power spectrum $C^{ac}_{\ell''}$ above may be approximated as $C^{ac}_{\ell''}\simeq(C^{ac}_\ell+C^{ac}_{\ell'})/2\equiv C^{ac}_{(\ell,\ell')}$. Combined with the following useful property of the map-level mode-coupling coefficients
    \begin{align}
      \sum_{\ell''m''}{\cal M}_{\ell m,\ell''m''}(w_a)\,{\cal M}^{*}_{\ell' m',\ell''m''}(w_c) \notag \\ ={\cal M}_{\ell m,\ell'm'}(w_aw_c),
    \end{align}
    this then allows us to approximate the Gaussian covariance as:
    \begin{align}\nonumber
      {\rm Cov}(\tilde{C}^{ab}_\ell,\tilde{C}^{cd}_{\ell'})= \;
      &C^{ac}_{(\ell,\ell')}C^{bd}_{(\ell,\ell')}\Xi_{\ell\ell'}(w_aw_c,w_bw_d)\\\label{eq:nka_simple}
      &+(c\leftrightarrow d).
    \end{align}
    Calculating the Gaussian covariance thus reduces to estimating the same set of coupling coefficients as the pseudo-$C_\ell$ estimator, this time involving the real-space product of pairs of masks. Additionally, we require a reasonable estimate of the true underlying power spectra of the fields involved (which could potentially be approximated from the measured spectra themselves).

    As described in \cite{2010.09717}, the accuracy of the NKA covariance can be further improved in two ways. The first important effect is the residual impact of mode coupling in the power spectrum before it is pulled out of the sum in Eq. \ref{eq:general_2pt}. This is important, particularly for complicated masks, where the core assumption of the NKA (the narrowness of the map-level kernels) is less accurate. This may be achieved by replacing the true power spectra in Eq. \ref{eq:nka_simple} (e.g., $C_\ell^{ac}$) by
    \begin{equation}\label{eq:clbar0}
      \overline{C}^{ab}_\ell\equiv\frac{\left\langle \tilde{C}^{ac}_\ell\right\rangle}{\langle w_aw_c\rangle_{\rm sky}},
    \end{equation}
    where the numerator is the mode-coupled spectrum (Eq. \ref{eq:pcl}), and the denominator is the sky average of the product of both masks
    \begin{equation}
      \langle w_aw_c\rangle_{\rm sky}\equiv\int\frac{d\nv}{4\pi}\,w_a(\nv)w_c(\nv).
    \end{equation}
    Here, $\langle\tilde{C}^{ab}_\ell\rangle$ may be calculated from a guessed theory $C_\ell$ by convolving it with the pseudo-$C_\ell$ mode-coupling matrix, or it may simply be replaced by the pseudo-$C_\ell$ measured directly from the data.

    Secondly, all cosmological datasets are affected by some form of measurement noise (e.g., shape noise in cosmic shear), the properties of which (e.g., its variance) may vary across the sky. Approximating this noise to be uncorrelated between different pixels, we can calculate its impact on the general two-point correlators needed to calculate the Gaussian covariance (Eq. \ref{eq:general_2pt}) as
    \begin{align}\label{eq:general_2pt_wnoise}
      \langle\tilde{a}_{\ell m}\tilde{c}_{\ell'm'}^*\rangle
      =&\sum_{\ell''m''}{\cal M}_{\ell m,\ell''m''}(w_a)\,{\cal M}^{*}_{\ell' m',\ell''m''}(w_c)C_{\ell''}^{ac}\\\nonumber
      &+\delta_{ac}^K\,{\cal M}_{\ell m,\ell'm'}\left(\Omega_{\rm pix}\sigma_a^2w_a^2\right),
    \end{align}
    where the map $\sigma_a^2(\nv)$ denotes the noise variance in the pixel along direction $\nv$, and $\Omega_{\rm pix}$ is the pixel area. This additional contribution leads to the introduction of ``signal-signal'', ``signal-noise'', and ``noise-noise'' components of the total covariance matrix (see Eq. 2.29 of \cite{2010.09717}), all of which depend on pseudo-$C_\ell$ mode-coupling coefficients involving either the fields' masks or their mask-weighted noise variance maps.

    This improved NKA (iNKA) has been found to be remarkably accurate even for data with highly complex sky masks and inhomogeneous noise properties.

  \subsection{Catalogue-based \texorpdfstring{$C_\ell$}{CL}s}\label{ssec:pcls.cat}
      \begin{figure}
          \centering
          \includegraphics[width=0.49\textwidth]{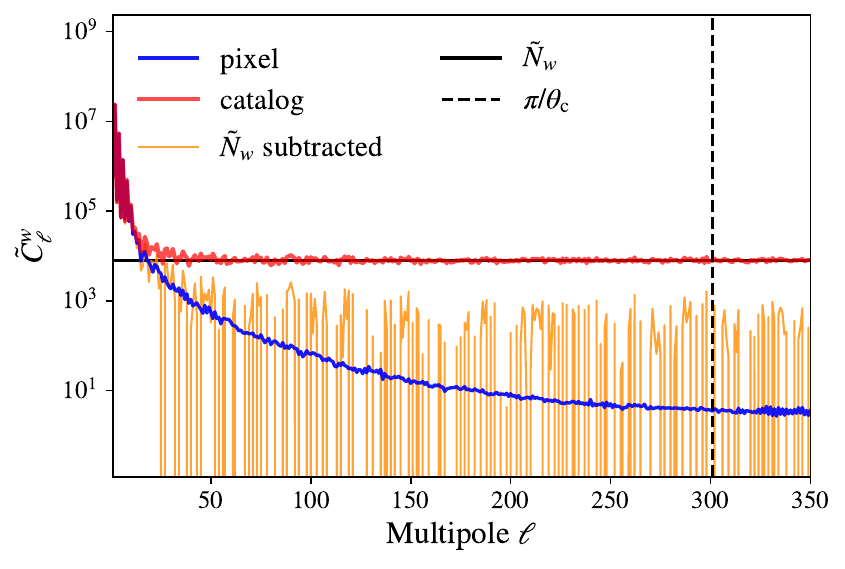}
          \caption{Mask pseudo power spectrum for the {\it Quaia} survey mask, comparing catalogue- and mask-based estimators. At the catalogue level, the constant shot noise floor ($\tilde{N}_w$) is reached at scales far below the mean interparticle scale ($\theta_{\rm c}$). For the catalogue estimates, we Poisson-sample $10^5$ positions with number densities following the survey mask and compute the spherical harmonics at the catalogue level.}
          \label{fig:Nw}
    \end{figure}

    Numerous cosmological datasets rely on measuring quantities of astrophysical interest at the discrete positions of observed sources. Canonical examples of this are cosmic shear, based on the observed ellipticity of galaxies; fast radio bursts (FRBs), based on the observed dispersion measure; and galaxy clustering, where the galaxy density is defined by the discrete positions of galaxies. In general, such a catalogue-based field may be described in terms of its mask $w_a(\nv)$ and the masked map $\tilde{a}(\nv)$ as:
    \begin{align}\label{eq:mask_delta}
       &w_a(\nv)\equiv\sum_iw_i\,\delta^D(\nv,\nv_i),\\ \label{eq:masked_cat_map}
       &\tilde{a}(\nv)=\sum_iw_ia_i\,\delta^D(\nv,\nv_i).
    \end{align}
    Here, all sums run over all sources in the catalogue, $w_i$ is the weight associated with the $i$-th source, $\nv_i$ are its sky coordinates, $a_i$ is the field value measured at that position, and $\delta^D(\nv_1,\nv_2)$ is the Dirac delta function defined on the sphere. The spherical harmonic transforms of the mask and masked map are then
    \begin{equation}
      w^a_{\ell m}=\sum_iw_i Y_{\ell m,i}^*,\hspace{12pt}
      \tilde{a}_{\ell m}=\sum_iw_ia_iY_{\ell m,i}^*,
    \end{equation}
    with $Y_{\ell m,i}\equiv Y_{\ell m}(\nv_i)$. The introduction of efficient algorithms for the calculation of discrete SHTs on irregular grids \cite{2304.10431} makes it possible to calculate these spherical harmonic coefficients directly from the catalogues, circumventing the need to bin sources into regular finite pixel grids, and the numerical instabilities (e.g., aliasing, finite-resolution effects) that come with it. This may, in fact, lead to a computational speed-up relative to the standard (map-based) pseudo-$C_\ell$ approach when targeting relatively large scales or sparse samples.

    As described in \cite{2407.21013}, the only additional step in estimating power spectra for catalogue-based fields is to account for shot noise in the power spectra of both the mask and the masked field for auto-correlations. Consider, for example, the mask power spectrum $\tilde{C}_\ell^w$: in the case of a continuous and reasonably simple mask, this power spectrum attains its largest values at low $\ell$, with the monopole $\ell=0$ being proportional to the total sky area covered, and then quickly drops and oscillates around zero. This decay is typically fast enough that one need only consider the mask power spectrum up to a modest maximum multipole $\ell_{\rm max}$, typically much lower than the multipole scale of the mean particle separation. If needed, the oscillatory behaviour may be suppressed by apodising the mask. In the case of a catalogue-based mask, where sources are randomly sampled over a similar sky area, the power spectrum is also largest at low $\ell$s, again tracking the total area covered by the sample, but then drops towards a constant shot-noise value $\tilde{N}_w$, given by the contribution from self-pairs to the power spectrum:
    \begin{equation}\label{eq:Nw}
      \tilde{N}_w\equiv\frac{1}{4\pi}\sum_iw_i^2.
    \end{equation}
    We show this explicitly in Fig.~\ref{fig:Nw}. Likewise, the power spectrum of the masked field reaches a noise-like floor $\tilde{N}_a$, given by the contribution from self-pairs:
    \begin{equation}\label{eq:Na}
      \tilde{N}_a\equiv\frac{1}{4\pi}\sum_iw_i^2a_i^2.
    \end{equation}
    As shown in \cite{2407.21013,2408.16903}, subtracting these noise-like components from the mask and the masked field auto-spectra yields a numerically stable and unbiased estimator that is also completely immune to any white-noise component (homogeneous or otherwise).

    At this point, the challenges in adapting the iNKA approach to compute covariances of catalogue-based power spectra become apparent: the approximation involves calculating power spectra of map products (hidden inside the coupling coefficients in Eq. \ref{eq:nka_simple}), as well as their sky average in Eq. \ref{eq:clbar0}. For catalogue-based masks (Eq. \ref{eq:mask_delta}), such products are either zero, in the case of cross-correlations between different catalogues, or they diverge for auto-correlations (since they involve products of coincident delta functions). Regularising this behaviour is the main topic of this paper.

\section{Analytical covariances for catalogue-based fields}\label{sec:theo}
  \subsection{Disjoint catalogues and cross-correlations}\label{ssec:theo.alldiff}
    For simplicity, let us start by considering a covariance matrix element of the form ${\rm Cov}(\tilde{C}^{ab}_\ell,\tilde{C}^{cd}_{\ell'})$, in which all fields involved, including their source catalogues, are different. Consider first the sky average $\langle w_aw_c\rangle_{\rm sky}$ required by the iNKA (Eq. \ref{eq:clbar0}). A priori, the product of the two masks is zero, and thus so should $\langle w_aw_b\rangle_{\rm sky}$. Note, however, that we may write this quantity in terms of the pseudo-$C_\ell$ of both masks, using Parseval's theorem:
    \begin{equation}
      \langle w_aw_b\rangle_{\rm sky}=\sum_{\ell=0}^{\ell_{\rm max}}\frac{2\ell+1}{4\pi}\tilde{C}^{w_a,w_b}_\ell.
    \end{equation}
    As we argued in the last section, the $\tilde{C}^{w_aw_a}_\ell$ generally peaks at $\ell=0$, scales with the area over which both catalogues overlap, then decays, eventually oscillating around zero. The sum above should therefore be dominated by the lowest multipoles, and it only reaches zero in the limit $\ell_{\rm max}=\infty$ after all the incoherent oscillations in $\tilde{C}^{w_aw_b}_\ell$, amplified by the prefactor $2\ell+1$, cancel out. A similar expression appears when considering the numerator of Eq. \ref{eq:clbar0}:
    \begin{equation}
      \left\langle\tilde{C}^{ab}_\ell\right\rangle=\sum_{\ell''=0}^{\ell_{\rm max}}\frac{2\ell''+1}{4\pi}\tilde{C}^{w_aw_b}_{\ell''}\,\Upsilon^{ab}_{\ell\ell''},
    \end{equation}
    where
    \begin{equation}
      \Upsilon^{ab}_{\ell\ell''}\equiv\sum_{\ell'=0}^{\ell{\rm max}}(2\ell'+1)\,C^{ab}_{\ell'}\,\wtj{\ell}{\ell'}{\ell''}{0}{0}{0}^2.
    \end{equation}
    \begin{figure}
        \centering
        \includegraphics[width=0.49\textwidth]{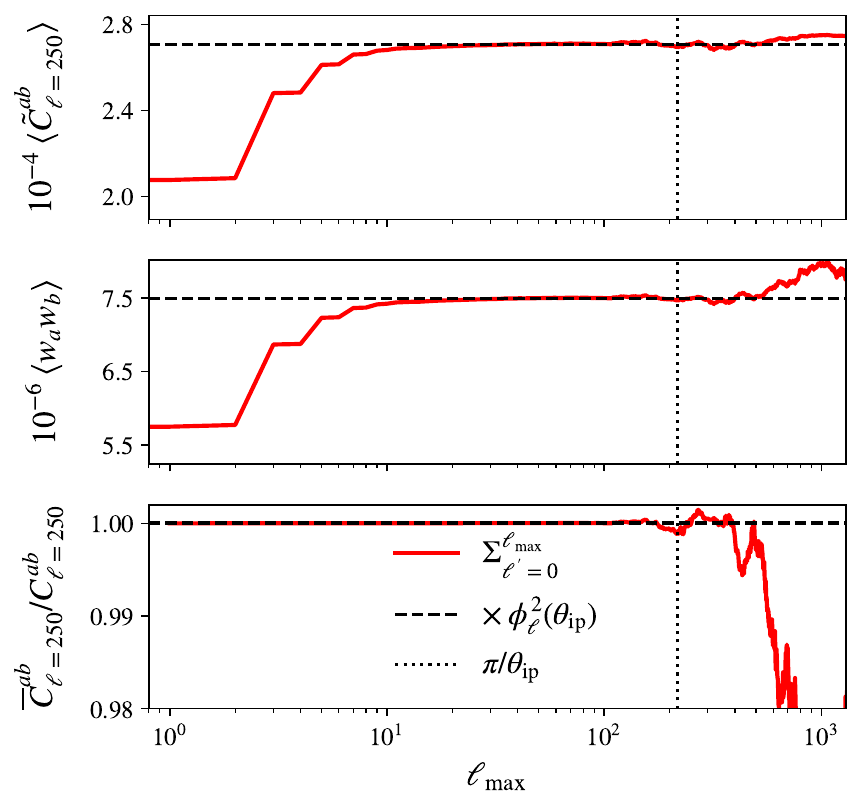}
        \caption{The ingredients used to calculate the effective power spectrum used in the iNKA approach (see Eq. \ref{eq:inka_cell}) as a function of the $\ell_{\rm max}$ used, for a simulated tracer catalogue. The last row shows the ratio between the iNKA spectrum and the input theory $C_\ell$. The calculation converges quickly and remains numerically stable on scales $\ell_{\rm max}\lesssim \pi/\theta_{\rm ip}$, where $\theta_{\rm ip}$ is the mean interparticle distance of the catalogue (vertical dotted line). The dashed line shows the result of treating each catalogue source as a Gaussian ``cloud'' with size $\theta_{\rm ip}$. These results are for a realisation of a sampled scalar field generated as described in Section \ref{ssec:val.general}, containing $3\times10^4$ sources.}
        \label{fig:inka_lmax}
    \end{figure}
    Although both expressions are strictly correct only in the limit $\ell_{\rm max}\rightarrow\infty$, we can consider their behaviour as a function of $\ell_{\rm max}$. This is shown in Fig. \ref{fig:inka_lmax} for one of the simulated catalogues used in Section \ref{ssec:val.general}. As the figure shows, both quantities seemingly converge at $\ell_{\rm max}\ll\ell_{\rm ip}\equiv\pi/\theta_{\rm ip}$, where $\theta_{\rm ip}$ is the mean inter-source distance in the sample. Interestingly, their ratio is remarkably stable, even at very low values of $\ell_{\rm max}$. For values of $\ell_{\rm max}\gtrsim\ell_{\rm ip}$, the sums become more numerically unstable.

    Choosing a finite value for $\ell_{\rm max}$ (e.g., $\ell_{\rm ip}$) is equivalent to band-limiting the harmonic transform of the mask or, in other words, treating each source not as a point but as a ``cloud'' of finite size $\theta_{\rm c}$. For the purposes of covariance matrix estimation, we therefore begin by applying Gaussian smoothing to the mask's harmonic coefficients with standard deviation $\theta_{\rm c}$. I.e. we replace:
    \begin{align}\label{eq:wbar}
      w^a_{\ell m}\rightarrow \bar{w}^a_{\ell m}\equiv w^a_{\ell m}\phi^a_\ell,\hspace{6pt}
      \phi^a_\ell\equiv\exp\left[-\frac{\ell(\ell+1)\theta_{\rm c}^2}{2}\right].
    \end{align}
    In order to account for both the finite interparticle distance of a given catalogue, as well as the possible band limit $\ell_{\rm max}$ imposed when analysing a given field, we use $\theta_c=\sqrt{\theta_{\rm ip}^2+(\pi/\ell_{\rm max})^2}$. Fig. \ref{fig:inka_lmax} shows that this choice reproduces well the numerically stable value to which the sums above converge. Crucially, replacing the delta function with Gaussian kernels addresses the other key problem with the iNKA approach when applied to catalogues. The resulting smoothed masks are defined continuously across the sky, allowing us to compute their real-space product in a numerically stable manner.

    From this discussion, we may now posit the following expression for the covariance of power spectra involving distinct catalogue-based fields:
    \begin{align}\nonumber
      {\rm Cov}(\tilde{C}^{ab}_\ell,\tilde{C}^{cd}_{\ell'})=
      &\overline{C}^{ac}_{(\ell,\ell')}\overline{C}^{bd}_{(\ell,\ell')}\Xi_{\ell\ell'}(\bar{w}_a\bar{w}_c,\bar{w}_b\bar{w}_d)\\\label{eq:nka_catdiff}
      &+(c\leftrightarrow d),
    \end{align}
    where $\bar{w}_a$ is the smooth mask given by Eq. \ref{eq:wbar}, and
    \begin{align}\nonumber
      &\overline{C}^{ab}_\ell\equiv\frac{\left\langle\tilde{C}^{ab}_\ell\right\rangle}{\langle \bar{w}_a\bar{w}_b\rangle_{\rm sky}},\\
      &\langle\bar{w}_a\bar{w}_b\rangle_{\rm sky}=\sum_{\ell}\frac{2\ell+1}{4\pi}\tilde{C}^{w_aw_b}_\ell\phi^a_\ell\phi^b_\ell. \label{eq:inka_cell}
    \end{align}

  \subsection{Auto-correlations}\label{ssec:theo.allsame}
    The case of covariance matrix elements involving more than one instance of the same catalogue requires additional care. On the one hand, the shot-noise contribution to the pseudo-$C_\ell$ of the masks and the masked map, arising from self-pairs, may lead to numerical instabilities. On the other hand, this shot noise contributes to statistical uncertainty and must therefore be accounted for when calculating the covariance matrix.

    We start by considering the general two-point correlator $\langle\tilde{a}_{\ell m}\tilde{c}^*_{\ell'm'}\rangle$, first introduced in Eq. \ref{eq:general_2pt} as the building block of the Gaussian covariance. For an auto-correlation, we can separate this into the contribution from distinct source pairs and self-pairs:
    \begin{align}\nonumber
      \langle \tilde{a}_{\ell m}\tilde{a}^*_{\ell'm'}\rangle
      =&\sum_{i\neq j}w_iw_j\langle a_ia_j\rangle Y^*_{\ell m,i}Y_{\ell'm',j}\\\nonumber
      &+\sum_i w_i^2\langle a_i^2\rangle\,Y^*_{\ell m,i}Y_{\ell'm',i}\\\label{eq:general_2pt_cat}
      =&\sum_{i\neq j}(\cdots)+{\cal M}_{\ell m,\ell'm'}(\overline{(aw_a)^2}).
    \end{align}
    Here, the last term is the map-level coupling coefficient (defined in Eq. \ref{eq:map_mcm}) for a map given by
    \begin{equation}\label{eq:self_var}
      \overline{(aw_a)^2}(\nv)\equiv\sum_i w^2_i\langle a_i^2\rangle\,\delta^D(\nv,\nv_i),
    \end{equation}
    which tracks the weighted variance of $a$ at the position of each source. Eq. \ref{eq:general_2pt_cat} is strongly reminiscent of the expression for the two-point correlator of a field with white noise, introduced in Eq. \ref{eq:general_2pt_wnoise}, with the self-pairs contribution playing the role of noise, and the sum over distinct pairs that of the signal.

    We can then estimate the contribution from different pairs (i.e. the first term in the previous equation) following the same rationale we used in the case of different-catalogue covariances: we construct a mask by replacing each source with a Gaussian cloud, and use the square of this mask to construct the mode-coupling coefficients $\Xi_{\ell\ell'}$. In doing so, we must avoid contributions to the square mask from self-pairs, since these are included in the second term of Eq. \ref{eq:general_2pt_cat}. To calculate this contribution, let us write the harmonic coefficients of the square mask, separating out the contribution from self-pairs:
    \begin{align}
      (\bar{w}^2)_{\ell m}
      &=\sum_{i\neq j}(\cdots)+\sum_i w_i^2\,(\phi^2_a)_\ell\,Y^*_{\ell m,i},
    \end{align}
    where $(\phi^2_a)_\ell$ is the harmonic transform of the square of $\phi_a$:
    \begin{align}
      (\phi^2_a)_\ell
      &\equiv 2\pi\int_{-1}^1 d\mu\,\phi^2_a(\mu)\,P_\ell(\mu)\\
      &=\frac{1}{4\pi\theta_{{\rm c},a}^2}\exp\left[-\frac{1}{4}\ell(\ell+1)\theta_{{\rm c},a}^2\right],
    \end{align}
    for a Gaussian kernel, where $P_\ell(\mu)$ are the Legendre polynomials. We can thus write the contribution to the squared mask from self-pairs as another catalogue-based field, given by:
    \begin{equation}
      (\overline{w^2})_{\ell m}\equiv \sum_i w_i^2(\phi^2_a)_\ell\,Y^*_{\ell m,i}.
    \end{equation}

    With this, we can then write the iNKA version of the two-point correlator in Eq. \ref{eq:general_2pt} as
    \begin{align}\label{eq:general_2pt_cat_inka}
      \langle \tilde{a}_{\ell m}\tilde{a}^*_{\ell'm'}\rangle & \simeq \overline{C}^{aa}_{(\ell,\ell')}{\cal M}_{\ell m,\ell'm'}(\bar{w}^2-\overline{w^2}) \notag \\ & \quad +{\cal M}_{\ell m,\ell'm'}(\overline{(aw)^2}).
    \end{align}
    Here, we must remember to subtract all self-pair contributions to $\overline{C}^{aa}_\ell$, since they are already included in the second term. That is, we define
    \begin{align}
      &\overline{C}^{aa}_\ell\equiv\frac{\tilde{C}^{aa}_\ell-\tilde{N}_a}{\langle \bar{w}^2-\overline{w^2}\rangle},\\
      &\langle \bar{w}^2-\overline{w^2}\rangle=\sum_\ell\frac{2\ell+1}{4\pi}(\tilde{C}^{ww}_\ell-\tilde{N}_w)(\phi^a_\ell)^2,
    \end{align}
    with $\tilde{N}_a$ and $\tilde{N}_w$ given in Eqs. \ref{eq:Na} and \ref{eq:Nw}, respectively. As we noted earlier, the structure of Eq. \ref{eq:general_2pt_cat_inka} is similar to that of a field with inhomogeneous white noise, as described in \cite{2010.09717}, with the self-pair contribution serving as the inhomogeneous noise component. We can thus calculate the power spectrum covariance for catalogue-based fields by including signal-noise and noise-noise terms that account for the shot noise contribution from self-pairs.

    It is important to note that the parallel between the structure described here and the signal/noise contributions described in \cite{2010.09717} is more than a simple coincidence: the noise-like contribution from self-pairs is an intrinsic aspect of fields sampled at discrete points that penalises sparser samples with larger statistical uncertainties.

  \subsection{General covariance structure}\label{ssec:theo.general}
    Putting together the results presented in the preceding two sections, we can write a general expression for the covariance matrix of pseudo-$C_\ell$s of catalogue-based fields:
    \begin{widetext}
    \begin{align}\nonumber
      {\rm Cov}(\tilde{C}^{ab}_\ell,\tilde{C}^{cd}_{\ell'})
      =&\,\overline{C}^{ac}_{(\ell,\ell')}\overline{C}^{bd}_{(\ell,\ell')}\Xi_{\ell\ell'}\left(\bar{w}_a\bar{w}_c-\delta^K_{ac}\overline{w_a^2},\bar{w}_b\bar{w}_d-\delta^K_{bd}\overline{w_b^2}\right)+\delta^K_{bd}\,\overline{C}^{ac}_{(\ell,\ell')}\Xi_{\ell\ell'}\left(\bar{w}_a\bar{w}_c-\delta^K_{ac}\overline{w_a^2},\overline{(bw_b)^2}\right)\\\label{eq:main_result}
      &+\delta^K_{ac}\,\overline{C}^{bd}_{(\ell,\ell')}\Xi_{\ell\ell'}\left(\overline{(aw_a)^2},\bar{w}_b\bar{w}_d-\delta^K_{bd}\overline{w_b^2}\right)+\delta^K_{ac}\,\delta^K_{bd}\Xi_{\ell\ell'}\left(\overline{(aw_a)^2},\overline{(bw_b)^2}\right)\\\nonumber
      &+(c\leftrightarrow d),
    \end{align}
    \end{widetext}
    where the different ingredients entering this expression are:
    \begin{align}\label{eq:inka_cell_final}
      &\overline{C}^{ab}_\ell\equiv\frac{\tilde{C}^{ab}_\ell-\delta_{ab}\tilde{N}_a}{\sum_{\ell'}\frac{2\ell'+1}{4\pi}(\tilde{C}^{w_aw_b}_{\ell'}-\delta_{ab}\tilde{N}_w)\phi^a_{\ell'}\phi^b_{\ell'}},\\
      &(\bar{w}_a)_{\ell m}=\sum_iw^a_i\phi^a_\ell\,Y^*_{\ell m,i},\\
      &(\overline{w_a^2})_{\ell m}\equiv\sum_i(w^a_i)^2(\phi^2_a)_\ell Y^*_{\ell m,i},\\\label{eq:awa2}
      &\left(\overline{(aw_a)^2}\right)_{\ell m}\equiv\sum_i(a_iw^a_i)^2Y_{\ell m,i}^*,\\
      &\phi^a_\ell\equiv e^{-\frac{1}{2}\ell(\ell+1)\theta^2_{{\rm c},a}},
      \hspace{12pt}(\phi^2_a)_\ell\equiv\frac{e^{-\frac{1}{4}\ell(\ell+1)\theta^2_{{\rm c},a}}}{4\pi\theta^2_{{\rm c},a}}.
    \end{align}
    $\tilde{N}_w$ and $\tilde{N}_a$ are given in Eqs. \ref{eq:Nw} and \ref{eq:Na}, respectively, and the coupling coefficients $\Xi_{\ell\ell'}(m_1,m_2)$ are defined in Eq. \ref{eq:Xi}. This is the main result of this paper.

    A final note concerns the calculation of the noise-noise component of the Gaussian covariance (the last term in Eq. \ref{eq:main_result}). This term involves constructing the local variance map in Eq. \ref{eq:self_var}, which requires an estimate of the field variance at each source. This is typically not immediately available and must be estimated directly from the data. A na\"ive ansatz, which we use in Eq.~\ref{eq:awa2}, is to simply replace $\langle a_i^2\rangle\rightarrow a_i^2$ for each source. However, the noise-noise contribution requires computing the pseudo-$C_\ell$ for this map, which in turn involves the fourth-order moment of $a_i$ under this approximation. In other words, the fact that $\langle a_i^2\rangle\langle a_j^2\rangle\neq\langle a_i^2a_j^2\rangle$ leads to an overestimation of the power spectrum of the local variance map at high $\ell$, which we must correct for. Under the assumption of Gaussian statistics, this can be done analytically using Wick's theorem (for more details, see Appendix \ref{app:awa2_shotnoise}). 
    
    Furthermore, to simplify the discussion, the derivation above has assumed that we are interested in the covariance matrix of the pseudo-spectrum $\tilde{C}^{ab}_\ell$. In reality, as we discussed in Section \ref{ssec:pcls.cat}, we must always subtract the contribution from self-pairs $\tilde{N}_a$ (Eq. \ref{eq:Na}) in catalogue-based fields to ensure the estimator is unbiased and numerically stable. Thus, more precisely, we are in fact interested in the covariance of the combination $\tilde{C}^{ab}_\ell-\delta_{ab}^K\tilde{N}_a$, which leads to a similar correction to the pseudo-$C_\ell$ of the local variance map. In summary, any occurrence of the auto-spectrum of $\overline{(aw_a)^2}$ in the calculation must have the following additive constant offset subtracted:
    \begin{equation}\label{eq:4point_corr_spin0}
      \tilde{N}_{\overline{(aw_a)^2}}\equiv\frac{1}{4\pi}\sum_iw_i^4a_i^4.
    \end{equation}
    Note that, just like $\tilde{N}_a$ is the contribution to the the auto-correlation $\tilde{C}^a_\ell$ from self-pairs, so is $\tilde{N}_{\overline{(aw_a)^2}}$ the contribution to the pseudo-spectrum of $\overline{(aw_a)^2}$ from self-pairs. A proof of this can be found in Appendix \ref{app:awa2_shotnoise}.

  \subsection{Momentum fields and galaxy clustering}\label{ssec:theo.clust}
    There is a subclass of catalogue-based fields, called ``momentum'' fields in this paper, that represent cosmological tracers weighted by the local galaxy density. An example is the radial momentum field $\pi_r\sim (1+\delta_g)\,v_r$, used in measuring the low-redshift kinematic Sunyaev-Zel'dovich effect \cite{2512.14625,2604.19744}, where the quantity weighted by the local number density is the spin-0 radial velocity field. Momentum fields can also be constructed from quantities with arbitrary spin and can be exploited, e.g., in the study of the moving-lens effect or in general directional stacking analyses \cite{2605.15947,2605.18938}. In these cases, the masked field takes the same form as that of the standard sampled field, Eq. \ref{eq:masked_cat_map}, where $a_i$ is the complex value of a spin-$s$ field at the position of the $i$-th source. However, unlike sampled fields, the mask of a momentum field represents the \emph{expected} mean number density of sources (i.e., the number that would have been observed in an otherwise homogeneous universe), rather than the actual observed density. This is often called the ``completeness'' or ``depth'' map and may be represented either as a continuous map or as a set of random points that track the local expected source density. In the first case (a continuous mask), Eq.~\ref{eq:main_result} is modified by replacing the signal-signal covariance kernel with its standard iNKA expression, since shot noise is not an issue. 

    In turn, a mask based on randoms takes the form
    \begin{equation}
      w(\nv)=\sum_{r\in R} \alpha\,w_r\,\delta^D(\nv,\nv_r),
    \end{equation}
    where we have specified that the sum runs over the random points. The normalisation constant $\alpha\equiv\sum_{d=D}w_d/\sum_{r=R}w_r$ ensures that $w$ represents the local density of sources in the data. For simplicity of notation in what follows, we will absorb $\alpha$ into the random weights $w_i$ (i.e. they are normalised so that $\sum_{r\in R}w_r=\sum_{d\in D}w_d$).
    
    Galaxy clustering can itself be treated as a momentum field with the following minimal modifications:
    \begin{enumerate}
      \item Each source in the data is assigned a field value of unity.
      \item The mean density (estimated from the mask) is subtracted from the masked field.
    \end{enumerate}
    This implies, in this case, that the masked field is
    \begin{align}\nonumber
      \delta_g(\nv)
      &=\sum_{d\in D}w_d\,\delta^D(\nv,\nv_d)-w(\nv)\\\nonumber
      &=\sum_{d\in D}w_d\,\delta^D(\nv,\nv_d)-\sum_{r\in R}w_r\,\delta^D(\nv,\nv_r),
    \end{align}
    where, in the second line, we have assumed that the mask is defined in terms of randoms. Since the random catalogue also contributes to the shot noise of the masked map (i.e., the contribution from self-pairs), the noise terms in the covariance matrix must also include this contribution. That is, the map used to calculate the noise-like coupling coefficients in the case of galaxy clustering should be
    \begin{equation}
      \overline{(aw_a)^2}(\nv)=\sum_{d\in D}w_d^2\,\delta^D(\nv,\nv_d)+\sum_{r\in R}w_r^2\,\delta^D(\nv,\nv_r),
    \end{equation}
    instead of Eq. \ref{eq:self_var}. Likewise, for galaxy clustering, the shot-noise contribution to the pseudo-spectrum must account for randoms,
    \begin{equation}
      \tilde{N}_a=\frac{1}{4\pi}\left(\sum_{d\in D}w_d^2+\sum_{r\in R}w_r^2\right)\, .
    \end{equation}
    Finally, no additive correction to the noise-noise covariance $\Delta\tilde{C}^{\overline{(aw_a)^2}}_\ell$ (Eq.~\ref{eq:4point_corr_spin0}) is needed, as there is no fourth-order term $\langle a_i^2 a_j^2 \rangle$ in the case of clustering. These modifications do not apply to other types of momentum fields, in which the mean density contribution from randoms is not subtracted from the masked map. However, in any momentum field using randoms, all mask elements entering the Gaussian covariance in Eq. \ref{eq:main_result} (e.g., $\bar{w}_a$, $\overline{w_a^2}$, $\tilde{N}_w$, $\tilde{C}^{w_aw_b}_\ell$), as well as the four-point additive correction (Eq.~\ref{eq:4point_corr_spin0}) must be calculated from the randoms, and not the data.

  \subsection{Software implementation}\label{ssec:theo.nmt}
    The covariance estimator described above is implemented in the public code \nmt\footnote{\url{https://github.com/LSSTDESC/NaMaster}}. While the user-facing interface for estimating covariances is unchanged from previous versions of the code (e.g., as presented in \cite{1906.11765}), the code includes several enhancements to facilitate the seamless treatment of catalogue-based fields. In particular, \nmt supports three new types of field objects representing such quantities:
    \begin{itemize}
      \item \texttt{NmtFieldCatalog} objects represent fields with general spin sampled at the positions of discrete sources, where the positions of these sources constitute the field's effective mask. Observables such as cosmic shear or FRB dispersion measure should be represented by these objects.
      \item \texttt{NmtFieldCatalogMomentum} objects represent ``momentum'' fields as described in Section \ref{ssec:theo.clust}, in which a field of arbitrary spin is sampled at the discrete positions of sources, but the field's mask is represented by either a pixelated depth/completeness map or a random catalogue. These objects may be used to represent, for example, the reconstructed galaxy momentum, which is used in measurements of the kSZ and moving-lens effects \cite{2512.14625,2605.15947,2605.18938}.
      \item \texttt{NmtFieldCatalogClustering} objects represent the source overdensity. As described in Section \ref{ssec:theo.clust}, they are effectively a subclass of momentum fields, in which the sampled field is simply 1 everywhere.
    \end{itemize}
    These catalogue-based field objects support many of the same features as standard \texttt{NmtField} objects, including linear deprojection of systematic templates, as described in \cite{2510.19912}. New functionality has also been implemented to seamlessly calculate the effective power spectra $\overline{C}^{ab}_\ell$ used by the iNKA approach (Eq. \ref{eq:inka_cell_final}). Further information and examples may be found in the \nmt documentation pages\footnote{\url{https://namaster.readthedocs.io}}.

  \subsection{Exact summation}\label{ssec:theo.brute}
For sufficiently sparse samples, it is feasible to evaluate the covariance of the decoupled power spectra exactly, without resorting to the iNKA. We describe this calculation here for the auto-correlation of a single catalogue-based scalar field $a$. The decoupled estimator $\hat{C}_\ell^{aa}$ is obtained from the pseudo-$C_\ell$ (Eq.~\ref{eq:pcl_def}) by subtracting the shot-noise bias $\tilde{N}_a$ (Eq.~\ref{eq:Na}) and applying the inverse of the mode-coupling matrix (Eq.~\ref{eq:pcl}),
\begin{equation}\label{eq:invert}
    \hat{C}_{\ell}^{aa} = \sum_{\ell_1} [M^{-1}]_{\ell \ell_1} \, \big(\tilde{C}_{\ell_1}^{aa} - \tilde{N}_a\big).
\end{equation}
Inserting this into the covariance and expanding the product gives
\begin{equation} \label{eq:cov2}
\begin{split}
\mathrm{Cov}&(\hat{C}_\ell^{aa}, \hat{C}_{\ell'}^{aa}) =  \sum_{\ell_1,\ell_2} [M^{-1}]_{\ell\ell_1} [M^{-1}]_{\ell_2\ell'} \\ &\times  \Big[ \left\langle \tilde{C}_{\ell_1}^{aa} \tilde{C}_{\ell_2}^{aa} \right\rangle - \left\langle \tilde{N}_a\,\tilde{C}_{\ell_1}^{aa} \right\rangle 
- \left\langle \tilde{N}_a\,\tilde{C}_{\ell_2}^{aa} \right\rangle + \left\langle \tilde{N}_a^2\right\rangle\Big] \\ & - \left\langle \hat{C}_\ell^{aa} \right\rangle \left\langle \hat{C}_{\ell'}^{aa} \right\rangle,
\end{split}
\end{equation}
where $\ell_1$ and $\ell_2$ are summed multipoles.
 
Under the assumption of Gaussianity, each correlator in Eq.~\ref{eq:cov2} reduces by Wick's theorem to products of the field two-point function evaluated at pairs of sources. Splitting the field value at the $i$-th source into signal and measurement-noise contributions, $a_i = s_i + n_i$, this two-point function reads
\begin{equation} \label{eq:field_corr}
    \left\langle a_i\,a_j\right\rangle = \left\langle s_i\,s_j\right\rangle + \sigma_{N}^2\, \delta^K_{ij},
\end{equation}
with uncorrelated noise $\langle n_i n_j\rangle = \sigma_{N}^2 \delta^K_{ij}$, and per-source signal variance
\begin{equation}
    \sigma_{S}^{2} \equiv \langle s_{i}^2\rangle = \sum_\ell \frac{2\ell + 1}{4\pi}\, \hat{C}_\ell^{aa}.
\end{equation}
The noise variance $\sigma_{N}^2$ is fixed by contracting both sides of Eq.~\ref{eq:field_corr}.
 
Carrying out the sums over $m$ with the spherical-harmonic addition theorem,
\begin{equation}
  \sum_m Y_{\ell m, i}^* Y_{\ell m, j} =\frac{2\ell + 1}{4\pi}P_{\ell, ij},
\end{equation}
where $P_{\ell, ij} \equiv P_\ell(\nv_i \cdot \nv_j)$ are the Legendre polynomials, and using the fact that the estimator is unbiased, so that $\langle \hat{C}_\ell^{aa}\rangle\langle \hat{C}_{\ell'}^{aa}\rangle = C_\ell^{aa}\,C_{\ell'}^{aa}$, the covariance becomes
\begin{widetext}
\begin{equation} \label{eq:cov_brute_long}
\begin{split}
    \mathrm{Cov}(\hat{C}_\ell^{aa}, \hat{C}_{\ell'}^{aa}) = & \sum_{\ell_1,\ell_2} [M^{-1}]_{\ell\ell_1} [M^{-1}]_{\ell_2\ell'} \Bigg[\langle \tilde{C}_{\ell_1}^{aa} \rangle \langle\tilde{C}_{\ell_2}^{aa} \rangle
 \\ + & \sum_{i,j, p, q}  \frac{w_i w_jw_p w_q}{(4\pi)^2} P_{\ell_1, ij}   P_{\ell_2, pq} \left\langle a_i a_p \right\rangle \left\langle a_j a_q \right\rangle 
+ \sum_{i,j, p, q}  \frac{w_i w_j w_p w_q}{(4\pi)^2} P_{\ell_1, ij} P_{\ell_2, pq}
  \left\langle a_i a_q \right\rangle \left\langle a_j a_p \right\rangle\\
-&\; \langle \tilde{N}_a \rangle \langle\tilde{C}_{\ell_1}^{aa} \rangle-  \sum_{i, j, k} \frac{2 w_k^2 w_i\,w_j}{(4\pi)^2} P_{\ell_1, ij}
 \left\langle a_i\,a_k\right\rangle \left\langle a_j\,a_k\right\rangle 
- \langle \tilde{N}_a \rangle \langle\tilde{C}_{\ell_2}^{aa} \rangle-  \sum_{i, j, k} \frac{2 w_k^2 w_i\,w_j}{(4\pi)^2} P_{\ell_2, ij}
 \left\langle a_i\,a_k\right\rangle \left\langle a_j\,a_k\right\rangle \\
+& \left(\tilde{N}_{w} ( \sigma_{S}^{2} + \sigma_{N}^{2})\right)^2 +  \sum_{ij}  \frac{2 w_i^2 w_j^2 }{(4\pi)^2} \left\langle a_i\,a_j\right\rangle^{2} \Bigg]
- C_\ell^{aa}\,  C_{\ell'}^{aa}.
\end{split}
\end{equation}
\end{widetext}
 
This expression can be simplified considerably. The two signal--signal terms (the second and third) are identical under relabelling of the dummy indices. The expectation value of the pseudo-$C_\ell$ is the coupled signal spectrum plus the self-pair shot-noise floor, which splits into a field-variance and a measurement-noise contribution,
\begin{equation}
    \left\langle \tilde{C}_{\ell}^{aa} \right\rangle = \sum_{\ell_1} M_{\ell\ell_1} C_{\ell_1}^{aa}  + \tilde{N}_{w} \, \sigma_{S}^{2} + \tilde{N}_{w} \, \sigma_{N}^{2},
\end{equation}
with $\tilde{N}_w$ the mask shot noise (Eq.~\ref{eq:Nw}). Finally, separating each source sum into its equal- and distinct-index parts,
\begin{equation}
\begin{aligned}
    \sum_{i,j, p, q} &= \sum_{i=j, p= q} + \sum_{i=j, p\not= q} + \sum_{i\not=j, p= q} + \sum_{i\not=j, p\not= q},\\
    \sum_{i,j, k} &= \sum_{i=j, k} + \sum_{i\not=j, k},
\end{aligned}
\end{equation}
the equal-index (self-pair) contributions are exactly those removed by the debiasing in Eq.~\ref{eq:invert} (see also Appendix \ref{app:awa2_shotnoise}), while the remaining disconnected pieces cancel against $C_\ell^{aa}C_{\ell'}^{aa}$. The covariance, therefore, reduces to
\begin{equation} \label{eq:cov_brute}
\begin{split}
    \mathrm{Cov}&(\hat{C}_\ell^{aa}, \hat{C}_{\ell'}^{aa})  =  2\sum_{\ell_1,\ell_2} [M^{-1}]_{\ell\ell_1} [M^{-1}]_{\ell_2\ell'} \\
    &\times \sum_{i \not= j,\, p \not= q}\frac{w_i w_jw_p w_q}{(4\pi)^2}   P_{\ell_1, ij} P_{\ell_2, pq}
     \left\langle a_i a_p \right\rangle \left\langle a_j a_q \right\rangle.
\end{split}
\end{equation}
The equivalence of this expression with Eq.~\ref{eq:main_result} is shown in Appendix \ref{app:brute_nka}.

 Eq.~\ref{eq:cov_brute} is computationally prohibitive for large catalogues, as it scales with the number of sources $N_\mathrm{sources}$ and the maximum value of the multipoles $\ell_\mathrm{max}$  as $\mathcal{O}(N_{\mathrm{sources}}^4 \ell_\mathrm{max}^4)$. In addition, memory restrictions may apply depending on the available hardware. We therefore implement it using the \texttt{jax} \cite{jax2018github} library, with details given in Appendix \ref{app:brute_code}.

\section{Validation}\label{sec:val}
  We now turn to validating the analytical covariance formalism developed in Sect.~\ref{sec:theo} against sample covariance estimators using simulated data. We start with mock-data tests on several relevant analysis cases: sparse fast radio burst (FRB) catalogues with low field-level noise, dense cosmic-shear mock surveys with high field-level noise, and galaxy clustering mock surveys. In the last section, we move on to a larger suite of noiseless sky simulations covering many types of cross- and autocorrelation in a multi-tracer analysis, including catalogue-sampled fields, galaxy clustering, momentum fields, and regular (pixelated) maps.
  
  \subsection{Sparse catalogues: FRBs}\label{ssec:val.sparse}
    \begin{figure*}
        \centering
        \includegraphics[width=0.8\textwidth]{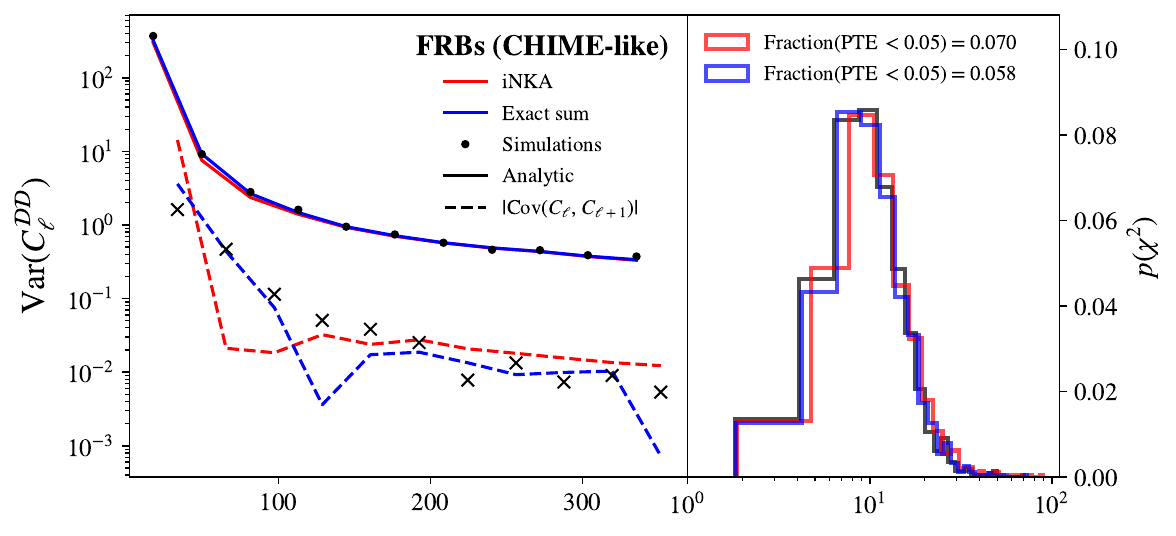}
        \includegraphics[width=0.8\textwidth]{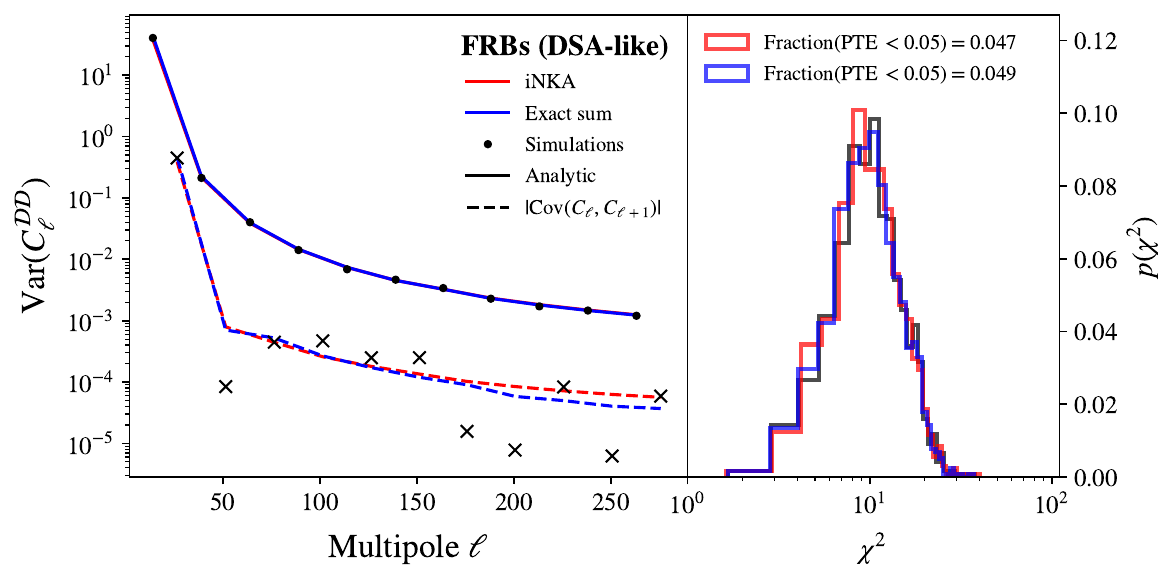}
        \caption{Covariances of dispersion-measure auto-power spectra for mock FRBs at 3615 CHIME positions (upper panel) and 50,000 mock catalogue positions for a DSA-like survey (lower panel). We compare analytical covariances using iNKA (red) and the exact summation method (blue) with empirical covariances (black) in the left-hand panels, and show the $\chi^2$ distribution from 1000 mock catalogues in the right-hand panels.}
        \label{fig:FRB}
    \end{figure*}
  
    A typical example of a very sparse tracer are FRBs, millisecond-long powerful radio pulses of unknown, likely extragalactic origin \cite{2007Sci...318..777L}. Currently, a few thousand FRB sources are known, with by far the largest number having been detected by the Canadian Hydrogen Intensity Mapping Experiment \citep[CHIME,][]{2026ApJS..283...34C}. While we expect this number to grow significantly over the next decade, the corresponding catalogues will remain sparse overall, with expected number densities $\bar{n}\sim 10^{-3}\;\mathrm{arcmin}^{-2}$. The FRB dispersion measure quantifies the frequency-dependent time delay of FRB photons as they propagate through plasma in the intervening interstellar and intergalactic media. Typical measured values are on the order of $\text{DM}_0\approx 1000\,\text{pc}\,\text{cm}^{-3}$. Therefore, owing to their low numbers, the limiting factors in measuring angular cross-correlations with FRBs are shot noise and the inhomogeneous sky footprint. As demonstrated in \cite{2407.21013}, both effects can be analytically controlled by using catalogue-based power spectrum estimators, which have been employed in recent cross-correlation analyses involving FRBs \cite{2506.08932,2607.04106}.

    Since FRB power spectra become shot-noise-dominated at relatively modest scales ($\ell\lesssim 100$), we can compare both iNKA and exact-summation covariance estimators with the fully empirical estimate at reasonable computational cost. We generate 1000 realisations of a Gaussian signal in harmonic space, with an angular power spectrum calculated as in \cite{2026JCAP...06..006R} using the redshift distribution inferred from the CHIME catalogue. To test our covariance estimator on the DM auto-power spectrum, we consider a realistic CHIME-like scenario with 3615 source positions provided by the CHIME Collaboration \cite{2026ApJS..283...34C}, and a futuristic scenario with $5\times 10^4$ sources, compatible with the sample that could be observed by the Deep Synoptic Array (DSA) \cite{2019BAAS...51g.255H,2025JCAP...12..035B}. The positions of these sources are Poisson-sampled non-uniformly according to the CHIME Collaboration's detection probability mask. For simplicity, every source has equal survey weight $w_i\equiv 1$. We construct mock catalogues as follows: to every source, we assign a DM by evaluating $a^{\rm DM}_{\ell m}$ at its exact position through an inverse spherical harmonic transform using the irregular-grid method of \cite{2304.10431} implemented in the {\tt ducc} library\footnote{\url{https://mtr.pages.mpcdf.de/ducc}}. We add random noise drawn from a Gaussian with zero mean and, in the CHIME case, a standard deviation of 150 $\text{pc}\,\text{cm}^{-3}$, to mimic the expected shot noise error due to the Milky Way and the host galaxy contributions, as well as the intrinsic field noise from cosmic variance. For DSA, we assume a total standard deviation of 50 $\text{pc}\,\text{cm}^{-3}$, representing a test in a highly signal-dominated regime. We do not assume any noise bias, which, in practice, would be estimated from the mean DM across all sources and subtracted from each source, yielding the same result as for Gaussian noise. From these mock catalogues, we compute the catalogue-based power spectrum using \texttt{NmtFieldCatalog} objects and 12 logarithmically spaced multipole bins between $\ell_{\rm min}=2$ and $\ell_{\rm max}=383$ for CHIME and $\ell_{\rm max} = 300$ for DSA. Our reference covariance is an empirical estimate from 1000 simulations. The iNKA covariance estimates the local variance map from a single realisation (Eq.~\ref{eq:awa2}) and the iNKA power spectrum $\overline{C}^{ab}_\ell$ (Eq.~\ref{eq:inka_cell_final}) using the mode-coupled theory spectrum to avoid realisation-dependent noise bias\footnote{Note that, in the real-data scenario, we do not have access to the true theory spectrum. In the absence of a sufficiently good guess, we may instead use $\left\langle \tilde{C}_\ell \right\rangle\approx\tilde{C_\ell}$ directly from the data. This is, in fact, the default fallback option supported by \nmt. The advantage of this approach is that the constructed covariance is free from bias arising from the use of an incorrect theoretical model to describe the signal. The main drawback is that, in doing so, we introduce realisation-dependent stochastic noise in the analytical covariance and that the measured signal and its covariance become correlated.}.  We obtain an alternative estimate using the exact summation method of Section \ref{ssec:theo.brute}, applying Eq.~\ref{eq:cov_brute_long}. In particular, we use the coupled-theory spectrum to replace $\langle \tilde{C}_\ell\rangle$ and to estimate the field two-point correlators $\langle a_i a_j\rangle$ and the self-pair noise floor (Eq.~\ref{eq:Na}) from a single realisation. We use a \texttt{jax}-based implementation to speed up the computation, as described in Appendix~\ref{app:brute_code}. 

    Figure \ref{fig:FRB} in the upper panel shows the results for the CHIME-like case. We compare the diagonal and first off-diagonal elements of the dispersion-measure auto-spectrum covariances among the empirical, exact-summation, and iNKA methods, and find good agreement among all three. Small biases relative to the reference are observed in the second iNKA multipole bin and in the fourth bin of the exact summation off-diagonal. We explicitly checked that, for the exact summation, this can be explained by stochastic noise in the data-based estimates of the correlators $\langle a_i a_j\rangle$, which is important given CHIME's small catalogue size. The Narrow Kernel Approximation, which may hold less well for very sparse masks, may be responsible for the bias observed in the first three CHIME bins in the iNKA case. The $\chi^2$ distributions obtained from each of these three covariances, shown in the panel on the right, are in remarkably good agreement, with excellent right-tail coverage: the 5$\%$ tail in the exact summation method is covered by 5.8\% of the simulations, which rises slightly to 7\% for the iNKA covariance. Figure \ref{fig:FRB} in the lower panel shows the results for the DSA-like scenario, with noticeable improvements, both visually and statistically: analytical 5$\%$ quantiles are covered by 4.7\% and 4.9\% for the exact summation and iNKA covariances, respectively, indicating an excellent match among all three methods. Although there is visible scatter in the off-diagonal elements of the empirical covariance, which reflects statistical noise from using 1000 simulations, there is no indication that this biases the validation or that running more simulations would not yield almost exact agreement between the direct summation and the sample covariance.

  \subsection{Dense catalogues: cosmic shear}\label{ssec:val.dense}
    \begin{figure*}
      \centering
      \includegraphics[width=0.8\textwidth]{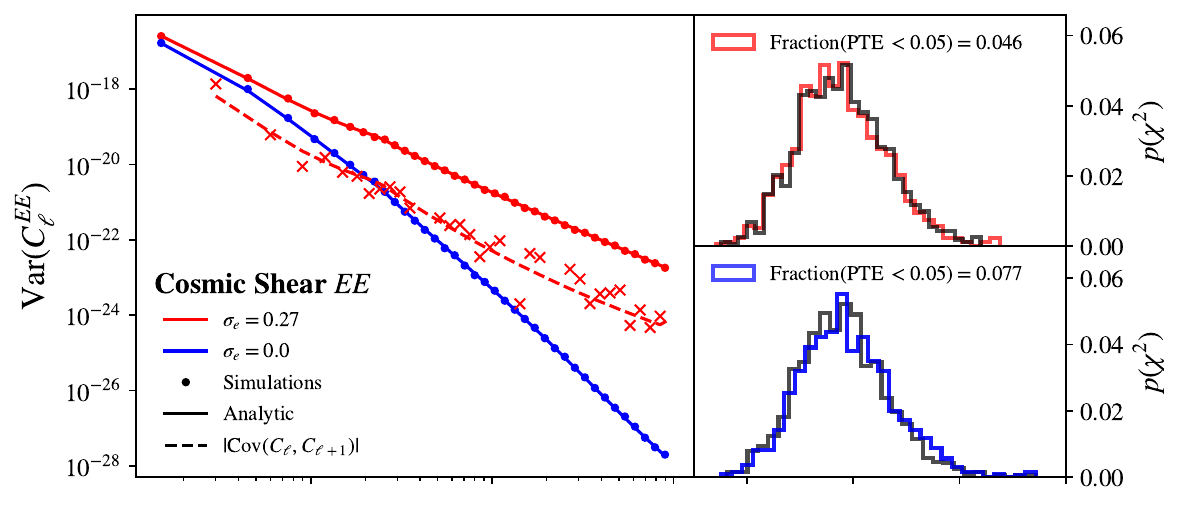}
      \includegraphics[width=0.8\textwidth]{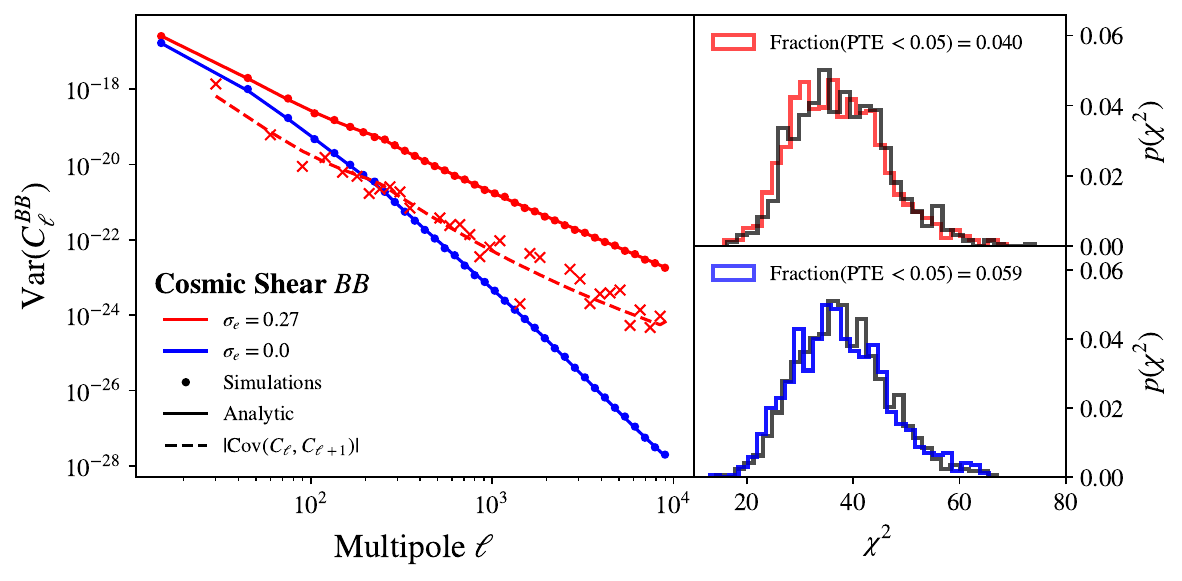}
      \caption{Power spectrum covariance for the DES-like simulated cosmic shear sample. Results are shown for the covariance between the $E$-mode and $B$-mode power spectra (left panels of the upper and lower figures). The right panels show the $\chi^2$ distributions calculated using the analytical covariance against the empirical reference distribution (black histograms). The blue lines, points, and histograms show results from simulations in which only the lensing shear signal is sampled at the source positions, whereas results including both the signal and Gaussian uncorrelated noise are shown in red. In the latter case, we assume a noise standard deviation per component of $\sigma_e=0.27$. The sample variance from 1000 realisations is shown as points, with the analytical prediction as solid lines. The dashed line and crosses show the first off-diagonal element of the covariance matrix.}
      \label{fig:cov_shear}
    \end{figure*}
    In this second example, we consider the regime opposite to that studied in the previous section: a high-density sample ($\bar{n}\sim O(1-10\,{\rm arcmin}^{-2})$) where power spectra can be recovered up to high multipoles, and where the field values sampled at any individual source are noise-dominated. These are the characteristics of cosmic shear data, a key cosmological observable for which catalogue-based power spectrum methods are particularly valuable \cite{2407.21013,2605.13543,2606.26223}. In this regime, a brute-force approach to the covariance matrix is computationally intractable, and the analytical approximations described here become most useful.

    To test our approach in a realistic setting, we use galaxy positions from the Dark Energy Survey (DES) Year-3 (Y3) cosmic shear sample \citep{2105.13543}. Specifically, we select galaxies from the third redshift bin used in the analysis of \cite{2403.13794}. We generate Gaussian realisations of the cosmic shear signal and evaluate them at the positions of these sources. These realisations were sampled from the theoretical $E$-mode power spectrum for the DES Y3 dataset, calculated assuming a {\sl Planck}-like cosmology \cite{1807.06209} using the Core Cosmology Library\footnote{\href{https://ccl.readthedocs.io/en/latest/index.html}{https://ccl.readthedocs.io}} \citep[CCL,][]{1812.05995}. The value of the spin-2 shear field at each source is then calculated using \texttt{ducc}. In addition to the cosmic shear signal, we add a shape noise component, uncorrelated between sources, with a standard deviation per component of $\sigma_e=0.27$. We will quantify the validity of our approximation both with and without this component to better assess the contributions of the different signal-like and noise-like mode-coupling coefficients in Eq. \ref{eq:main_result}.

    The sample consists of about $2.5\times10^7$ sources across approximately $4000\,{\rm deg}^2$, and the median interparticle distance is $\theta_{\rm ip}=0.15\,{\rm arcmin}$. Hence, power spectra can be reconstructed up to very large multipoles. To stress-test our approach, we measure power spectra up to $\ell_{\rm max}=10{,}000$. Although most cosmological shear analyses focus on larger scales \cite{2304.00701,2503.19441,2602.10065}, several works have shown that smaller, highly non-linear scales can be exploited with a sufficiently accurate and flexible parametrisation of non-linearities and baryonic effects \cite{2303.05537,2403.13794}. We measure the cosmic shear power spectrum in 1000 simulations using 38 bandpowers spanning multipoles $\ell\leq\ell_{\rm max}$. We use the same binning scheme adopted in \cite{2403.13794}, with linearly spaced bins up to $\ell=240$ and logarithmic spacing thereafter. The simulation measurements are then used to construct the sample covariance, which we compare with our analytical estimate. The analytical covariance is calculated following Eq. \ref{eq:main_result}, using the input power spectrum to estimate $\bar{C}^{ab}_\ell$, and with the weighted variance map (Eq. \ref{eq:awa2}) calculated from one of the simulations, chosen at random.

    The result of this exercise is shown in Fig. \ref{fig:cov_shear} for both $E$-mode and $B$-mode power spectra. We find that the analytical covariance accurately describes the sample covariance on all scales for $E$-modes and on scales $\ell\gtrsim10$ for $B$-modes. The off-diagonal elements of the covariance are small (a few per cent of the main diagonal) and are well characterised by the analytical estimate. Remarkably, these results also hold for noiseless simulations, where the contribution from the noise-like components of the covariance (which are exact in the analytical approach) is suppressed. In the panels on the right-hand side, we compare the empirical $\chi^2$ distribution with the iNKA result, finding very good visual agreement and right-tail statistical coverage: in the noisy case, the 5$\%$ analytical probability to exceed (PTE) is covered by 4$\%$ of the simulations, very slightly overestimating the covariance, and in the noiseless case there is a very mild tendency of underestimation (5.9$\%$ coverage). Overall, these results attest to an excellent match between the iNKA covariance and the reference.
    \begin{figure*}
      \centering
      \includegraphics[width=0.80\textwidth]{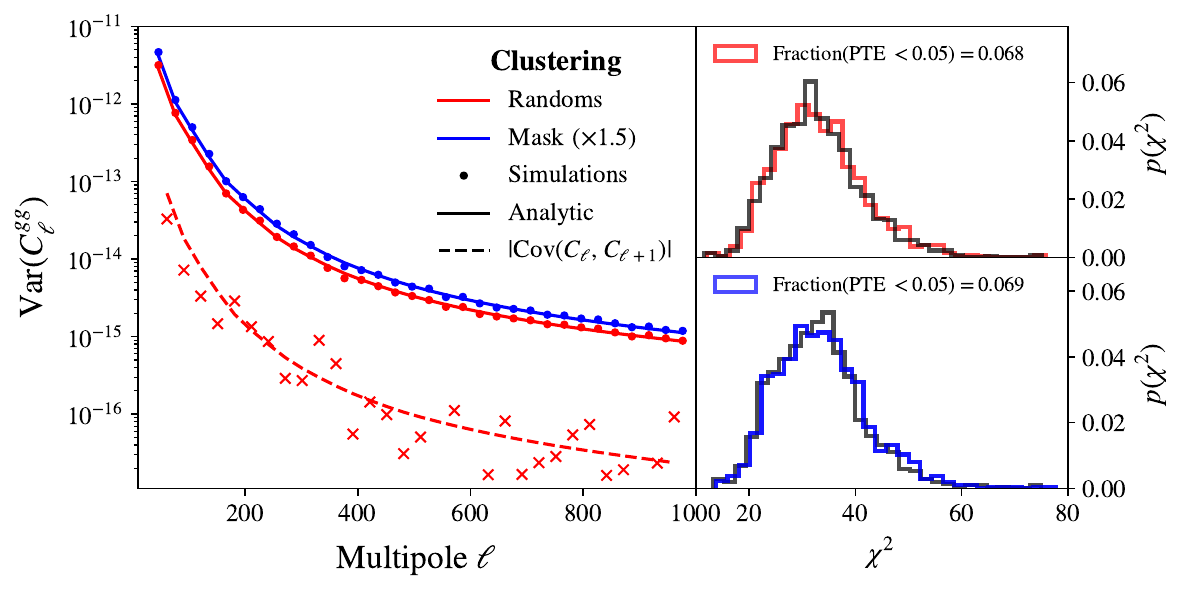}
      \caption{{\it Left panel:} Power spectrum covariance for the LRG-like simulated galaxy clustering sample. Results are shown for power spectra calculated using random catalogues to quantify the sample completeness (red) and using a continuous mask (blue). For clarity, the mask-based results have been rescaled by a factor of 1.5. The points show the sample variance estimated from 1000 realisations, with the analytical Gaussian covariance as solid lines. The dashed line and crosses show the first off-diagonal element of the covariance matrix. {\it Right panels:} corresponding analytical $\chi^2$ distributions from 1000 simulations, compared with the empirical distributions in black.}
      \label{fig:cov_lrgs}
    \end{figure*}

  \subsection{Galaxy clustering}\label{ssec:val.clust}
    In our final example, we explore galaxy clustering in a realistic setting. As discussed in Section \ref{ssec:theo.clust}, this is an important case to test separately, since the noise-like contributions to the covariance are driven by the Poisson-like shot noise of the sample rather than by the variance of any sampled field.

    As a realistic setup, we generate mock realisations of a galaxy clustering sample with properties (number density, clustering bias) compatible with those of the photometric luminous red galaxy (LRG) sample of the DESI Legacy Imaging Survey \cite{2309.06443,2407.04607} used in the analysis of \cite{2407.04607,2510.17796}. To generate mock realisations, we follow the steps outlined below:
    \begin{enumerate}
      \item We start by calculating the angular power spectrum of the sample. As before, we use CCL, assuming the redshift distribution for the second redshift bin from \cite{2407.04607} and a galaxy bias of $b=2$. We calculate this power spectrum up to $\ell_{\rm max}=1000$.
      \item We generate a Gaussian realisation of the galaxy overdensity $\delta_G$ from this power spectrum on a HEALPix \cite{astro-ph/0409513} map with resolution parameter $N_{\rm side}=512$.
      \item In order to generate a positive-definite overdensity map $\delta_g$ from which we can generate a Poisson-sampled catalogue, we perform a log-normal transformation \cite{1991MNRAS.248....1C,2111.05069,2302.01942} of the form
      \begin{equation}
        1+\delta_g=\exp[\delta_G-\sigma_G^2/2],
      \end{equation}
      where $\sigma_G^2\equiv\langle\delta_G^2\rangle$ is the variance of $\delta_G$.
      \item We generate a simulated catalogue via Poisson-sampling, where the probability of finding a galaxy in pixel $p$ is
      \begin{equation}
        p_p\propto w_p\,(1+\delta_{g,p}),
      \end{equation}
      where $w_p$ is a completeness map. The completeness map $w_p$ was constructed from the publicly available random catalogues used in \cite{2407.04607}.
    \end{enumerate}
    As in previous sections, we generate 1000 such simulations and estimate the power spectrum for each to compute the sample covariance. We calculate these power spectra in two ways: using the completeness map as a mask on the simulated catalogues, or using a random catalogue generated by Poisson sampling of this map. In the latter case, we use 10 times as many random sources as in the data catalogue.
    
    We note that the use of lognormal realisations in this case introduces a couple of subtleties to consider when interpreting the results. First, the simulated sampled overdensity field is not Gaussian, but results from a Cox process in which an underlying continuous field, sampled from a lognormal distribution, is Poisson-sampled. The lognormal and Poisson elements of the field introduce a non-zero connected trispectrum that is not captured by the Gaussian covariance (see, e.g., \cite{2507.03749} for an in-depth discussion of the Poisson trispectrum). These contributions are typically small, and we expect our Gaussian analytical covariance to still reproduce the empirical covariance with reasonable accuracy. Secondly, the lognormal transformation modifies the angular power spectrum of the resulting field relative to that of the Gaussian field (i.e., our theoretical power spectrum). Since the iNKA approach requires an estimate of the signal power spectrum, we measure this directly from the lognormal realisations before Poisson sampling, averaging over all 1000 realisations to suppress cosmic variance. Furthermore, since the pixelated lognormal field is then Poisson sampled, placing simulated sources at random positions within each pixel effectively smooths the signal on pixel scales. We must therefore account for the impact of the associated pixel window function on the theoretical power spectrum used in the iNKA covariance. We thus multiply the spectrum obtained from the 1000 lognormal realisations (which does not account for pixel smoothing) by the harmonic-space window function for \texttt{HEALPix} pixels with $N_{\rm side}=512$.

    The results of this comparison are shown in Fig. \ref{fig:cov_lrgs}. We find that the iNKA approach predicts the variance of the clustering power spectrum with high accuracy for the two cases explored here (using random catalogues and a continuous completeness map). The correlation structure is also accurately recovered, with the figure also showing the first off-diagonal elements of the covariance. We do not observe a significant impact of the connected trispectrum in the lognormal and Poisson distributions. An inspection of the correlation structure in the empirical covariance after subtracting the iNKA Gaussian prediction reveals only a tenuous component that is correlated across widely separated multipoles on small scales, perhaps consistent with expectations from Poisson statistics, although the amplitude of this component is too small to draw reliable conclusions. The $\chi^2$ distributions, shown in the panels on the right-hand side, show very good agreement between the empirical (black histograms) and the analytical iNKA estimate, both for randoms and mask. The right-tail coverage is good (6.8--6.9 $\%$), with a slight tendency of the analytical covariance to underestimate the true covariance, although safely negligible in practice. In summary, the iNKA estimator provides an accurate prediction of the covariance matrix of the galaxy-clustering catalogue-based power spectrum.

  \subsection{General method validation}\label{ssec:val.general}
    \begin{figure}
      \centering
      \includegraphics[width=0.49\textwidth]{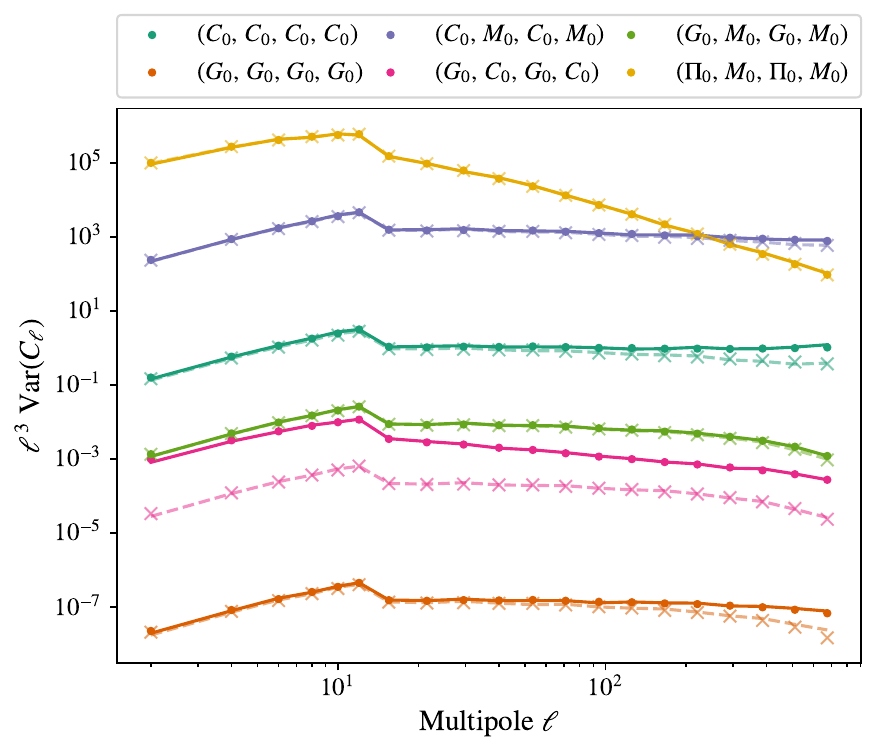}
      \caption{Power spectrum variance estimates for spin-0 field validation cases, comparing the iNKA approximation (lines) against the sample variance from 1000 simulations (markers). We show ``auto-catalogue'' covariances where all catalogues share their sources (solid lines, dots), or ``cross-catalogue '' covariances where the catalogues have no sources in common (dashed lines, crosses). The cases are offset by constant factors for visual purposes.}\label{fig:var_spin0}
    \end{figure}

    \begin{figure}
      \centering
      \includegraphics[width=0.49\textwidth]{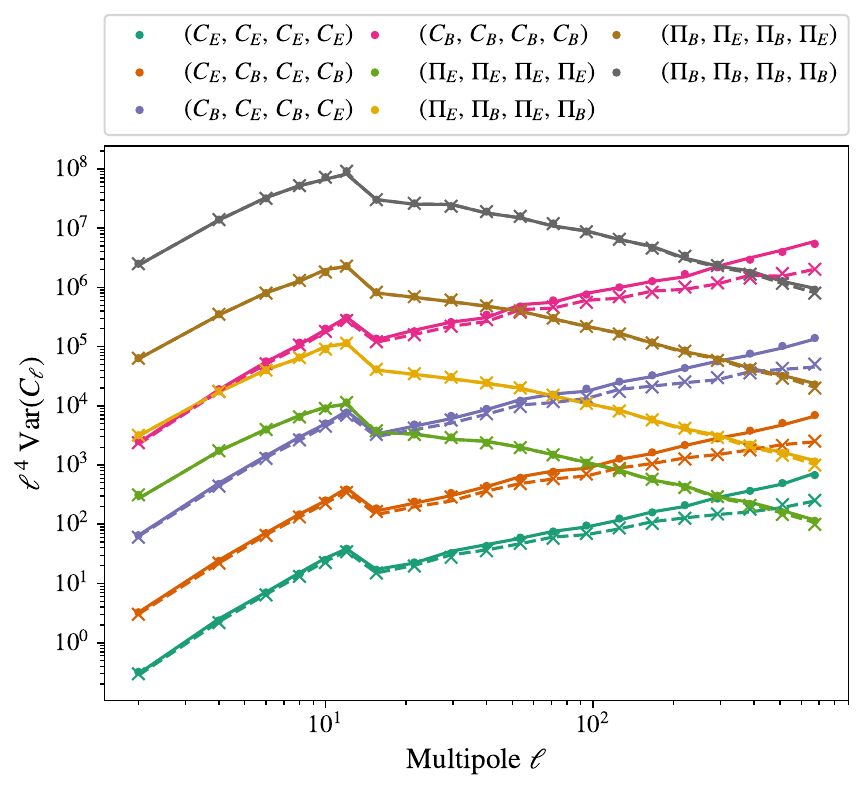}
      \caption{Same as Fig.~\ref{fig:var_spin0}, but showing the spin-2 validation cases.}\label{fig:var_spin2}
    \end{figure}

    \begin{figure}
      \centering
      \includegraphics[width=0.49\textwidth]{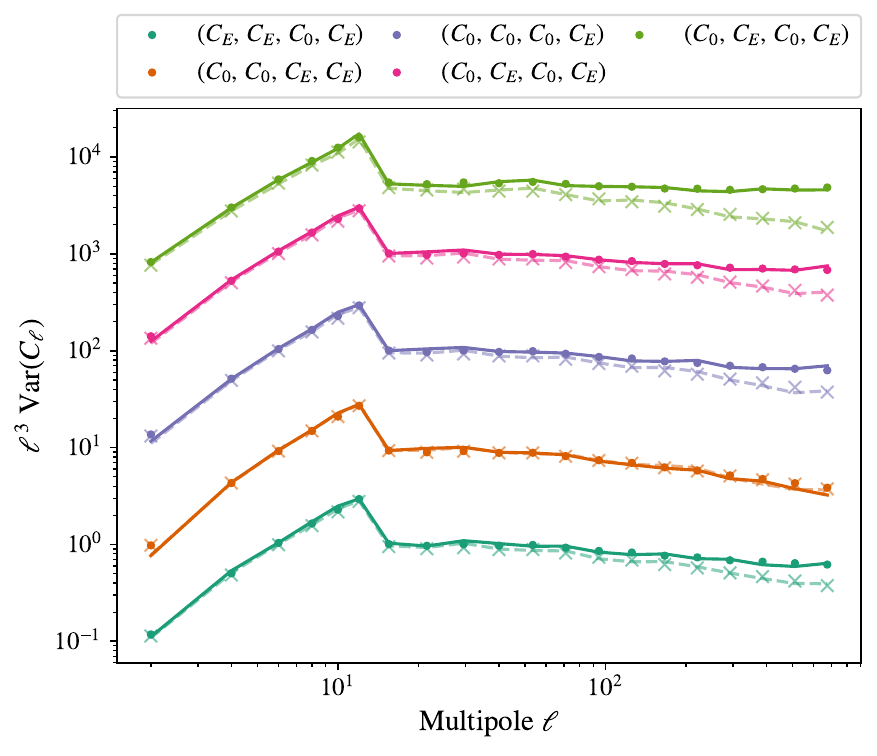}
      \caption{Same as Fig.~\ref{fig:var_spin0}, but showing the spin-0 $\times$ spin-2 validation cases.}\label{fig:var_spin2xspin0}
    \end{figure}
    
    \begin{figure*}
      \centering
      \includegraphics[width=\textwidth]{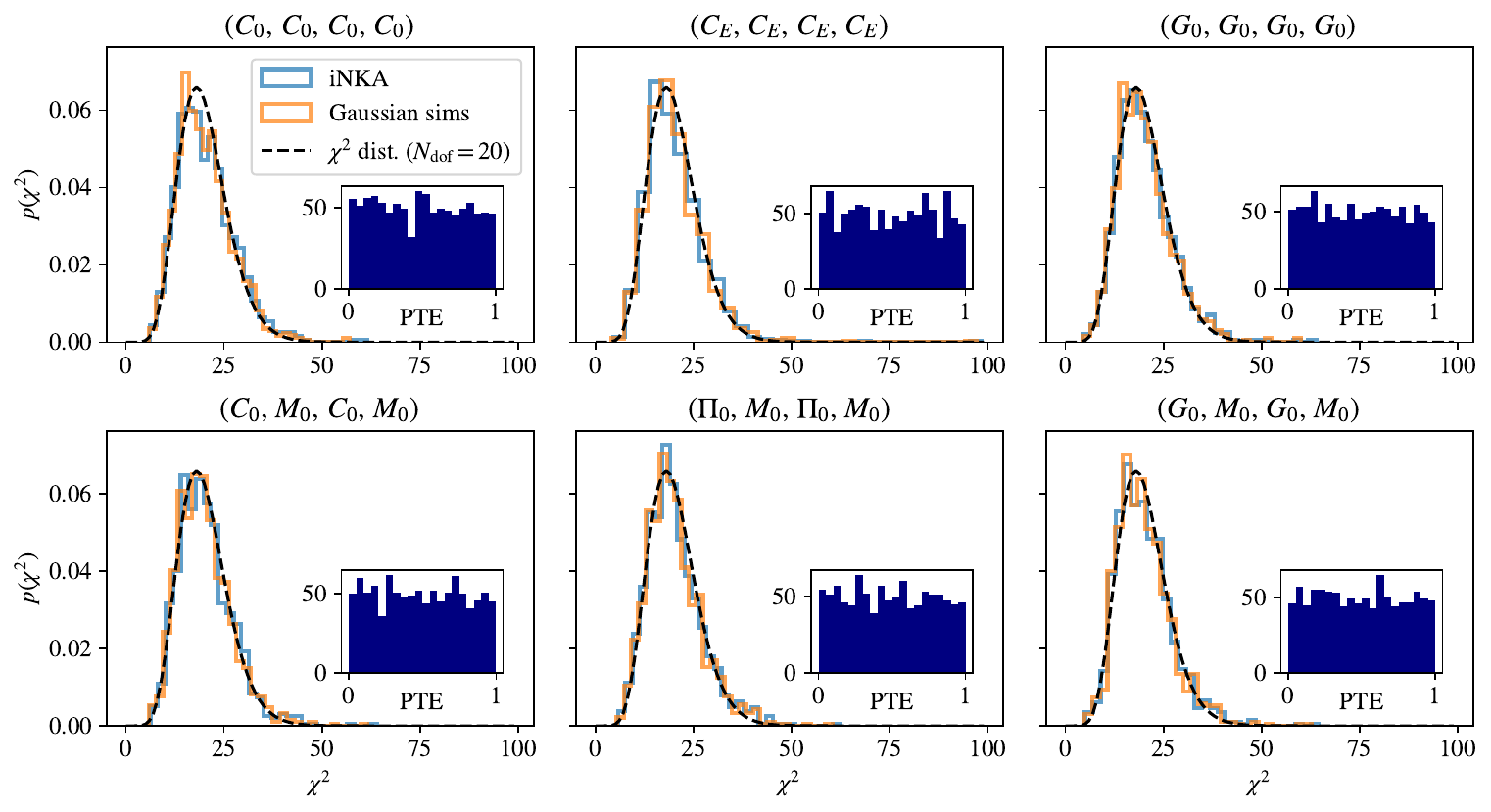}
      \caption{$\chi^2$ distributions for the iNKA and empirical covariances for a selection of validation cases, with all catalogue fields sharing the same sources. Small inset plots show the distribution of $\chi^2_{\rm iNKA}$ PTEs with respect to $\chi^2_{\rm sim}$.}\label{fig:chi2_overview}
    \end{figure*}
    In this section, we validate our method for general combinations of field types. As the four fields that form the legs of a given covariance matrix element, we consider catalogue-based fields of three different types:
    \begin{itemize}
      \item {\it Tracer catalogue.} These correspond to fields of arbitrary spin sampled at catalogue positions, which also constitute the mask. We draw the field's spherical harmonic coefficients as a Gaussian realisation of a power spectrum that follows a power-law form, $C_\ell\propto(10+\ell)^{-\alpha}$, with index $\alpha=0.7$. We tested different values of $\alpha$ and verified that the results presented here are insensitive to this choice. For spin-2 fields, we draw uncorrelated $E$- and $B$-mode realisations. We draw $10^6$ source positions as inhomogeneous Poisson realisations of an underlying completeness map $w_s$, for which we use the selection function from the {\sl Quaia} quasar catalogue \citep{2306.17749} at \texttt{HEALPix} resolution $N_{\rm side}=256$. The field is then evaluated at the catalogue positions using its spherical harmonic coefficients, with an irregular-grid inverse SHT with a maximum multipole $\ell_{\rm max}=767$. Finally, we use the field values and catalogue positions to create a \nmt \texttt{NmtFieldCatalog} object, assuming unit weights for simplicity, and then use it to estimate power spectra and covariances.
      \item {\it Galaxy clustering}. In this case, the catalogue positions represent the source overdensity we wish to analyse. To generate mock realisations we start by drawing a continuous overdensity map $\delta_g$ as a Gaussian realisation at resolution $N_{\rm side}=256$ from an input power spectrum $C_\ell^\delta \propto (10+\ell)^{-\alpha}$, where $\alpha=0.7$, and the proportionality factor that ensures the density $\propto1+\delta_g$ remains positive definite\footnote{More precisely, we set $C_\ell^\delta\equiv A_5\, C^0_\ell$, with $A_5^{-1}=5^2\sum_{\ell'=0}^{\ell_{\rm max}}(2\ell'+1)C_{\ell'}^0/(4\pi)$ and $\ell_{\rm max}=767$. This ensures that $\delta_g$ has a standard deviation of $1/5$, making unphysical crossings below $-1$ unlikely.}. We then draw source positions by Poisson sampling from a density map proportional to $w_s(1+\delta_g)$, where $w_s$ is the completeness map described above. We also generate random catalogues using the modulating mask $w_s$, which represents the sample's mean density and accounts for its inhomogeneous completeness. As indicated in Table~\ref{tab:validation}, the number of sources ranges from $10^6$ to $5\times10^7$, chosen to ensure that enough sources cover the noisy small scales of the underlying cosmological field, thereby making the empirical covariances numerically stable.\footnote{At high multipoles, the empirical covariance assumes its lowest values due to its strong dependence on the red input spectrum, which leads to Monte Carlo noise. This can be improved by choosing a less red spectrum and a higher number of sources. This choice does not affect the performance of the analytical covariance.} We use the resulting source positions to generate a \texttt{NmtFieldCatalogClustering} object, considering both fields where the completeness is represented by the pixelated $w_s$ map and by a random catalogue.
      \item {\it Momentum catalogue.} In this case, the field of interest is weighted by the local number density of sources. In this case, we generate source positions as described above for galaxy clustering, using $5\times 10^7$ sources. As in galaxy clustering, the number of sources is chosen so that the empirical covariance is sufficiently numerically stable to compare with the analytical covariance. We then compute the spherical harmonic coefficients $v_{\ell m}$ of a ``velocity'' field from a power spectrum $C_\ell^v=(10+\ell)^{-(\alpha+1)}$ and evaluate the field at the exact catalogue positions. We will explore cases where $v_{\ell m}$ is a scalar field, and where it is the $E$-mode of a spin-2 field. We then construct a \texttt{NmtFieldCatalogMomentum} object from the source positions, field values, and the positions of the random catalogue. We did not explore pixelated completeness masks for momentum fields due to their conceptual similarity to galaxy clustering. Finally, we obtain a theoretical prediction for the expected power spectrum of the momentum field $\pi=v\,(1+\delta)$ from $C_\ell^\delta$ and $C_\ell^v$ as described in Appendix A of \cite{2512.14625}, which we use to estimate the analytical covariance.
    \end{itemize}
    
    We compute the analytical covariance using the simulation ground-truth $C_\ell$s as input to the iNKA $\bar{C}_\ell^{ab}$ (Eq.~\ref{eq:inka_cell}), thereby avoiding realisation-dependent bias. Then we repeat the mock catalogue generation for 1000 different sky realisations, estimate their catalogue-based power spectra using the catalogue-based estimator \citep{2407.21013}, and compute their sample covariance. For the four fields that form the legs of the covariance, we may choose for some or all legs to be identical (``auto-catalogue'') or to differ in catalogue positions (``cross-catalogue''). Table \ref{tab:validation} lists the complete set of test cases, the spin modes, the number of sources and randoms, and the simulation input power spectrum. The last column (``KS PTE'') quantifies the agreement between the empirical and analytical covariance, as described below.

    \begin{table*}[t]
    \caption{\textbf{Validation cases for the binned iNKA covariance}, for the multipole range $\{2,\, 767\}$. The different columns are: the spin-$s$ field types that form the four legs of the covariance (tracer catalogue $C$, map $M$, galaxy clustering $G$, momentum field $\Pi$), their spin modes, the number of mock sources in the catalogue fields, the number of randoms (if randoms are being used for any of the 4 fields), the input power spectrum used to generate the sky modes, and the Kolmogorov-Smirnov (KS) PTEs assessing statistical agreement between empirical and analytical covariances for auto-catalogue and cross-catalogue scenarios.}
    \centering
    \scriptsize
    \setlength{\tabcolsep}{3pt}
    \renewcommand{\arraystretch}{1.12}
    \begin{tabular}{@{} l l l l l l p{0.06\textwidth} p{0.06\textwidth}@{}}
    \toprule
    \# &
    Field types
    & Spin modes
    & $N_{\rm sources}$
    & $N_{\rm randoms}$
    & Power spectrum
    & KS PTE (auto-catalogue)
    & KS PTE (cross-catalogue)\\
    \midrule

    1 &
    $(C_0,\,C_0,\, C_0,\, C_0)$
    & $(0,\, 0,\, 0,\, 0)$
    & $10^6$
    & -
    & $1/(10+\ell)^{0.7}$
    & 0.45 & 0.43 \\

    2 &
    $(C_0,\,M_0,\, C_0,\, M_0)$
    & $(0,\, 0,\, 0,\, 0)$
    & $10^6$
    & -
    & $1/(10+\ell)^{0.7}$
    & 0.71 & 0.97 \\

    3 &
    $(G_0,\,C_0,\, G_0,\, C_0)$
    & $(0,\, 0,\, 0,\, 0)$
    & $5\times 10^7$
    & $5\times 10^7$
    & $\propto\, 1/(10+\ell)^{0.7}$
    & 0.75 & 0.63\\

    4 &
    $(G_0,\,M_0,\, G_0,\, M_0)$
    & $(0,\, 0,\, 0,\, 0)$
    & $5\times10^7$
    & -
    & $\propto\, 1/(10+\ell)^{0.7}$
    & 0.94 & 0.19 \\

    5 &
    $(G_0,\,G_0,\, G_0,\, G_0)$
    & $(0,\,0,\, 0,\, 0)$
    & $5\times 10^7$
    & $5\times 10^7$
    & $\propto\, 1/(10+\ell)^{0.7}$  
    & 0.12 & \textcolor{red}{0.00013}$^\dagger$\\

    6 &
    $(\Pi_0,\,M_0,\, \Pi_0,\, M_0)$
    & $(0,\,0,\, 0,\, 0)$
    & $5\times 10^7$
    & $5\times 10^7$
    & $\delta_g$: $\propto 1/(10+\ell)^{0.7}$ 
    & 0.51 & 0.46 \\

    \midrule

    7 &
    $(C_2,\,C_2,\, C_2,\, C_2)$
    & $(E,\, E,\, E,\, E)$
    & $10^6$
    & -
    & $1/(10+\ell)^{0.7}$
    & 0.63 & 0.93 \\

    8 &
    $(C_2,\,C_2,\, C_2,\, C_2)$
    & $(E,\, B,\, E,\, B)$
    & $10^6$
    & -
    & $1/(10+\ell)^{0.7}$
    & 0.06 & 0.28 \\

    9 &
    $(C_2,\,C_2,\, C_2,\, C_2)$
    & $(B,\, B,\, B,\, B)$
    & $10^6$
    & -
    & $1/(10+\ell)^{0.7}$
    & 0.56 & 0.50 \\

    10 &
    $(\Pi_2,\,\Pi_2,\, \Pi_2,\, \Pi_2)$
    & $(E,\, E,\, E,\, E)$
    & $2\times 10^7$
    & $2\times 10^7$
    & $\delta_g$: $\propto\, 1/(10+\ell)^{0.7}$
    & 0.97 &  0.97 \\

    11 &
    $(\Pi_2,\,\Pi_2,\, \Pi_2,\, \Pi_2)$
    & $(E,\, B,\, E,\, B)$
    & $2\times10^7$
    & $2\times 10^7$
    & $\delta_g$: $\propto\, 1/(10+\ell)^{0.7}$
    & 0.81 & 0.38\\

    12 &
    $(\Pi_2,\,\Pi_2,\, \Pi_2,\, \Pi_2)$
    & $(B,\, B,\, B,\, B)$
    & $2\times 10^7$
    & $2\times 10^7$
    & $\delta_g$: $\propto\, 1/(10+\ell)^{0.7}$
    & 0.21 & 0.17\\

    \midrule

    13 &
    $(C_0,\,C_0,\, C_0,\, C_2)$
    & $(0,\, 0,\, 0,\, E)$
    & $10^6$
    & -
    & $1/(10+\ell)^{0.7}$
    & 0.81 & 0.54  \\

    14 &
    $(C_0,\,C_0,\, C_2,\, C_2)$
    & $(0,\, 0,\, E,\, E)$
    & $10^6$
    & -
    & $1/(10+\ell)^{0.7}$
    & 0.20 & 0.83 \\

    15 &
    $(C_0,\,C_2,\, C_0,\, C_2)$
    & $(0,\, E,\, 0,\, E)$
    & $10^6$
    & -
    & $1/(10+\ell)^{0.7}$
    & 0.51 & 0.71 \\

    16 &
    $(C_2,\,C_2,\, C_0,\, C_2)$
    & $(E,\, E,\, 0,\, E)$
    & $10^6$
    & -
    & $1/(10+\ell)^{0.7}$
    & 0.79 & 0.63 \\

    \bottomrule
    \end{tabular} \label{tab:validation} \\
    \footnotesize{$^\dagger$ The KS test passes if the multipole range $\ell>500$ is excluded.}
    \end{table*}

    As indicated in Table~\ref{tab:validation}, we separate the validation suite into three categories: Spin-0 auto- and cross-correlations of catalogues ($C_0$), galaxy clustering ($G_0$), momentum fields ($\Pi_0$), and standard pixelated maps ($M_0$,  cases 1--6), spin-2 auto-correlations of either tracer catalogue ($C_2$) or momentum fields ($\Pi_2$, cases 7--12) and, lastly, combinations for spin-0 and spin-2 tracer catalogue fields (cases 13--16). Figures \ref{fig:var_spin0}, \ref{fig:var_spin2}, and \ref{fig:var_spin2xspin0} show the variance of these field combinations across 1000 simulations, together with their iNKA predictions. In all cases explored, the iNKA analytical covariance reproduces the simulation-based estimate to a visually high accuracy.
    
    To make a more quantitative assessment of the accuracy of the catalogue-based iNKA approach, we compare the distributions of chi-squared values from both covariances. As before, we compute chi-squared statistics $\hat{\chi}^2_{M,i}$ for $M\in\{{\rm iNKA},\,{\rm sim}\}$ and $i$ running from 1 to 1000, with the empirical $\hat{\chi}^2_{\rm sim}$ considered the benchmark. For each case, we evaluate the per-simulation $\chi^2$ PTE values and compare the distribution across all simulations with the expected uniform distribution using a Kolmogorov-Smirnov (KS) test, as shown in the last two columns of Table~\ref{tab:validation}. We distinguish between ``cross-catalogue'' and ``auto-catalogue'' scenarios. Figure~\ref{fig:chi2_overview} shows the $\chi^2$ distributions for a representative subset of six validation cases in the ``auto-catalogue'' scenario. We compare empirical covariances (orange), iNKA analytical covariances (light blue), and a theoretical $\chi^2$ distribution with $n_{\rm bin}=20$ degrees of freedom, and show the $\hat{\chi}_{\rm iNKA}^2$ PTE distribution in the small inset plots.

    It is worth noting that, in the ``cross-catalogue'' case, the corresponding covariance matrix block is not a true covariance (i.e. it is not the full covariance matrix of a given data vector, but an off-diagonal block involving complementary elements of said vectors in either dimension). As such, it is not positive definite, and the resulting $\chi^2$ statistic is not expected to follow a chi-squared distribution. Furthermore, in the case of the simulation-based cross-catalogue covariance, the inverse matrix may be numerically unstable in noise-dominated regimes (e.g., high-$\ell$ in the presence of shot noise).
    
    The KS PTE values for the full set of 32 validation cases are shown in Table~\ref{tab:validation}. We note that, based solely on the number of tests performed, we expect a small proportion to exhibit PTEs outside the $(5,\,95)\%$ range. Almost all tests pass the KS test with ease, including both auto- and cross-catalogue cases. The only outlier is the cross-catalogue galaxy clustering test $(G_0,\, G_0,\, G_0,\, G_0)$ yielding a KS PTEs of $1.3\times 10^{-4}$. As discussed above, this is likely caused by Monte Carlo noise dominating the small scales in the empirical covariance, making the inverse covariance numerically unstable and thereby biasing the reference $\chi^2$ distribution. We verified that, by excluding the highest multipole bin at $\ell>500$, we obtain an acceptable PTE of 0.36.

\section{Conclusion}\label{sec:conc}
  We have presented an analytical approach for computing pseudo-$C_\ell$ covariances involving fields defined at the discrete source positions. The method is a generalisation of the improved Narrow Kernel Approximation (iNKA) of \cite{1906.11765,2010.09717}, addressing the challenge of computing the product of two masks when they are represented by discrete source positions. We have shown that this problem can be ``regularised'' by treating sources as clouds with a finite size commensurate with the interparticle distance in the source catalogue, and by separating the contributions to the covariance from self-pairs from those of different pairs. The self-pair contribution acts as an effective source of white noise and can be calculated exactly without approximations, whereas contributions from different pairs can be computed using the standard iNKA approach.

  We have shown that this approach provides an accurate estimate of the power spectrum covariance across a wide range of realistic setups, including cosmic shear and galaxy clustering from existing Stage-III experiments, as well as dispersion measure observations from sparse FRB catalogues. We have also carried out an exhaustive validation exercise, exploring all possible combinations of catalogue- and pixel-based fields with different spins, and found that the method performs well in all cases. Remarkably, this is even true in regimes where the standard NKA has been shown to struggle in past analyses, such as the $B$-mode power spectrum covariance in cosmic shear. This is likely due to the much more precise (in fact, exact) treatment of the effective noise due to self-pair correlations.

  Although we have shown that the method presented here remains accurate across a wide range of data types (sparseness, signal-to-noise ratio, spin, and $E/B$ asymmetry), it is worth noting some potential failure points. Firstly, although the noise-like contributions to the covariance due to self-pair correlations are exact when estimating the local noise variance from a large number of sources, the noise-like terms may exhibit additional scatter when estimated from extremely sparse catalogues. Additionally, the signal component involves invoking the Narrow Kernel Approximation, and thus will become increasingly inaccurate as:
  \begin{itemize}
    \item the signal power spectrum becomes steeper,
    \item the complexity of the mask (e.g., due to inhomogeneity in the source distribution) grows, or
    \item the signal exhibits increasing asymmetry between the $E$- and $B$-mode components for spin fields.
  \end{itemize}
  Therefore, the accuracy of this method may need to be quantified against Gaussian simulations when analysing data that falls within any of these categories, beyond the regimes tested here. Secondly, the iNKA has only been developed for fields with spins 0 and 2. This limitation arises because, as shown in \cite{1906.11765} (see also \cite{1609.09730}), the specific correlation structure for spin fields is derived by neglecting the spatial derivatives of the mask, and the number of derivatives depends on the spin. If an approximate covariance for fields with spins other than 0 or 2 is necessary, a potential alternative is to treat the $E$ and $B$-mode components of the field as independent scalar quantities. Although \cite{1906.11765} showed that this often yields a reasonable approximation to the covariance matrix, the accuracy is degraded relative to the full calculation in the spin-2 case. Therefore, this alternative approach should be tested before resorting to it.

  While these caveats should be borne in mind, we have shown that the extension presented here represents a significant improvement over the standard iNKA approximation, reproducing the disconnected covariance of a wide range of realistic datasets at high accuracy. This may be particularly useful in the context of multi-probe analyses in current and future LSS datasets, where the number of probes involved and the large cosmic volumes covered by the data make the use of large simulation ensembles impractical (or even unfeasible) for the purposes of covariance matrix estimation with the required level of accuracy and numerical stability.

\section*{Acknowldedgements}
  We would like to thank Nicolas Tessore and Tom Cornish for useful discussions. KW and DA are supported by the Science and Technology Facilities Council (STFC) under grants ST/X006344/1 and UKRI1164, and they acknowledge support from the Beecroft Trust. KW acknowledges the support of a Gianturco Junior Research Fellowship at Linacre College, University of Oxford. AN acknowledges support from the European Research Council (ERC) under the European Union’s Horizon 2020 research and innovation program with Grant agreement No. 101163128. EF acknowledges the support from the Deutscher Akademischer Austauschdienst (DAAD) Master Studies for All Academic Disciplines Scholarship Program. We made extensive use of computational resources at the University of Oxford Department of Physics, funded by the John Fell Oxford University Press Research Fund. We also made extensive use of the Marvin cluster at the University of Bonn.

  \emph{Software}:  We made extensive use of the {\tt numpy} \citep{oliphant2006guide, van2011numpy}, {\tt jax} \citep{jax2018github}, {\tt scipy} \citep{2020SciPy-NMeth}, {\tt astropy} \citep{1307.6212, 1801.02634}, {\tt healpy} \citep{Zonca2019}, {\tt matplotlib} \citep{2007CSE.....9...90H} and {\tt GetDist} \citep{Lewis:2019xzd} python packages, as well as the {\tt HEALPix} package \cite{astro-ph/0409513}. This paper makes use of software developed for the Large Synoptic Survey Telescope. We thank the LSST Project for making their code available as free software at \url{http://dm.lsst.org}. 

\onecolumngrid

\appendix
\section{Spin-2 fields}\label{app:spin_s}
  For simplicity, our discussion in Sections \ref{sec:pcls} and \ref{sec:theo} has focused on spin-0 fields. The generalisation of the results presented here to the case of spin-2 fields offers no major difficulty, beyond those already described in the literature. The modifications to the coupling coefficients appearing in Eq. \ref{eq:main_result} to accommodate different spins are described in \cite{1906.11765}, with the extensions to the noise-like contributions introduced in \cite{2010.09717}. The shot noise contribution to the field's pseudo-$C_\ell$, in the case of auto-spectra, as described in \cite{2407.21013}, is given by
  \begin{equation}
    \tilde{N}^{\alpha\beta}_a=\frac{\delta^{\alpha\beta}}{4\pi}\sum_iw_i^2\frac{a_{1,i}^2+a_{2,i}^2}{2},
  \end{equation}
  where $\alpha$ and $\beta$ label the different $E$ and $B$ components of the harmonic coefficients, and $a_{1/2,i}$ are the two different components of the field at the $i$-th source.

  The general covariance of spin-2 pseudo-$C_\ell$s must account for the polarisations ($E$ or $B$) of the different fields being correlated. The calculation is relatively straightforward, although it requires non-trivial bookkeeping. We reproduce the main results here for completeness.  As in the case of scalar fields, the first step is the calculation of a general two-point correlator of masked spin-2 fields (Eq. \ref{eq:general_2pt_cat}):
  \begin{equation}
    \langle\tilde{a}^\alpha_{\ell m}\tilde{a}^\beta_{\ell'm'}\rangle=\sum_{i\neq j}(\cdots)+\,_2{\cal M}^{\alpha\beta}_{\ell m,\ell'm'}(\overline{(aw_a)^2}),
  \end{equation}
  where the spin-2 map-level coupling coefficients are defined similarly to the spin-0 ones (Eq. \ref{eq:map_mcm}):
  \begin{equation}
    _2{\cal M}^{\alpha\beta}_{\ell m,\ell'm'}(w)\equiv\int d\nv\,w(\nv)\,\left(_2{\sf Y}^\dagger_{\ell m}(\nv)\,\,_2{\sf Y}_{\ell'm'}(\nv)\right)_{\alpha\beta},
  \end{equation}
  and $_2{\sf Y}_{\ell m}(\nv)$ are the spin-2 spherical harmonic matrices defined in Appendix A of \cite{1809.09603}. With this in hand and following the same approximations outlined in \cite{1906.11765}, we can calculate the final covariance matrix, including contributions from both self-pairs and different pairs. Concretely, the spin-$2$ version of Eq. \ref{eq:main_result} reads
  \begin{align}
    {\rm Cov}(\tilde{C}^{a_\alpha b_\beta}_\ell,\tilde{C}^{f_\phi g_\gamma}_{\ell'})=&\,\overline{C}^{a_{\alpha'}f_{\phi'}}_{(\ell,\ell')}\overline{C}^{b_{\beta'}g_{\gamma'}}_{(\ell,\ell')}\left[\Xi^{\alpha\phi,\alpha'\phi'}_{\beta\gamma,\beta'\gamma'}\right]_{\ell\ell'}\left(\bar{w}_a\bar{w}_f-\delta^K_{af}\overline{w_a^2},\bar{w}_b\bar{w}_g-\delta^K_{bg}\overline{w_b^2}\right)\\
    &+\delta^K_{bg}\,\overline{C}^{a_{\alpha'}f_{\phi'}}_{(\ell,\ell')}\left[\Xi^{\alpha\phi,\alpha'\phi'}_{\beta\gamma}\right]_{\ell\ell'}\left(\bar{w}_a\bar{w}_f-\delta^K_{af}\overline{w_a^2},\overline{(bw_b)^2}\right)\\
    &+\delta^K_{af}\,\overline{C}^{b_{\beta'}g_{\gamma'}}_{(\ell,\ell')}\left[\Xi^{\alpha\phi}_{\beta\gamma,\beta'\gamma'}\right]_{\ell\ell'}\left(\overline{(aw_a)^2},\bar{w}_b\bar{w}_g-\delta^K_{bg}\overline{w_b^2}\right)\\
    &+\delta^K_{af}\,\delta^K_{bg}\left[\Xi^{\alpha\phi}_{\beta\gamma}\right]_{\ell\ell'}\left(\overline{(aw_a)^2},\overline{(bw_b)^2}\right)\\\nonumber
      &+(c\leftrightarrow d).
  \end{align}
  Here, all repeated indices are implicitly summed over, and the spin-2 covariance coupling terms for the signal-noise ($\Xi^{\alpha\phi,\alpha'\phi'}_{\beta\gamma}$), noise-signal ($\Xi^{\alpha\phi}_{\beta\gamma,\beta'\gamma'}$), and noise-noise ($\Xi^{\alpha\phi}_{\beta\gamma}$) contributions are given by:
  \begin{align}
    &\left[\Xi^{EE}_{EE}\right]_{\ell\ell'}=
    \left[\Xi^{EE}_{BB}\right]_{\ell\ell'}=
    \left[\Xi^{BB}_{EE}\right]_{\ell\ell'}=
    \left[\Xi^{BB}_{BB}\right]_{\ell\ell'}=\,_2\Xi_{\ell\ell'}^+,\\
    &\left[\Xi^{EB}_{EB}\right]_{\ell\ell'}=
    \left[\Xi^{BE}_{BE}\right]_{\ell\ell'}=
    -\left[\Xi^{EB}_{BE}\right]_{\ell\ell'}=
    -\left[\Xi^{BE}_{EB}\right]_{\ell\ell'}=\,_2\Xi_{\ell\ell'}^-,\\
    &\left[\Xi^{EE}_{XY,XY}\right]_{\ell\ell'}=\left[\Xi^{BB}_{XY,XY}\right]_{\ell\ell'}=\,_2\Xi^+_{\ell\ell'},\\
    &\left[\Xi^{EB}_{EE,EB}\right]_{\ell\ell'}=\left[\Xi^{EB}_{BB,EB}\right]_{\ell\ell'}=\left[\Xi^{EB}_{BE,EE}\right]_{\ell\ell'}=\left[\Xi^{EB}_{BE,BB}\right]_{\ell\ell'}=\,_2\Xi^-_{\ell\ell'},\\
    &\left[\Xi^{BE}_{\beta\gamma,\beta'\gamma'}\right]_{\ell\ell'}=-\left[\Xi^{EB}_{\beta\gamma,\beta'\gamma'}\right]_{\ell\ell'}=\left[\Xi^{BE}_{\beta'\gamma',\beta\gamma}\right]_{\ell\ell'},\hspace{12pt}\left[\Xi^{\beta\gamma,\beta'\gamma'}_{\alpha\phi}\right]_{\ell\ell'}=\left[\Xi^{\alpha\phi}_{\beta\gamma,\beta'\gamma'}\right]_{\ell\ell'},
  \end{align}
  and the signal-signal terms ($\Xi^{\alpha\phi,\alpha'\phi'}_{\beta\gamma,\beta'\gamma'}$) are provided in Eqs. 2.63--2.65 of \cite{1906.11765}. Here, $_2\Xi^\pm_{\ell\ell'}$ are the spin-2 mode-coupling coefficients:
  \begin{equation}
    _2\Xi^{\pm}_{\ell\ell'}(v,w)=\sum_{\ell''}\frac{2\ell''+1}{4\pi}\tilde{C}^{v,w}_{\ell''}\frac{1\pm(-1)^{L}}{2}\wtj{\ell}{\ell'}{\ell''}{2}{-2}{0}^2,
  \end{equation}
  where $L\equiv\ell+\ell'+\ell''$.

  The only remaining piece is the procedure for computing the local weighted variance map, $\overline{(aw_a)^2}$ (see Eq. \ref{eq:self_var}), which is used to compute the coupling coefficients for the noise-like contributions to the covariance. As we discussed at the end of Section \ref{ssec:theo.general}, in the case of spin-0 fields, the only possible estimate of the local variance is $\langle a_i^2\rangle\simeq a_i^2$, in the case of spin-2 fields, we can follow from the definition of $\tilde{N}_a$ above and use $\langle a_i^2\rangle\simeq(a_{1,i}^2+a_{2,i}^2)/2$ as a local variance estimate. I.e. in the case of spin-2 fields, Eq. \ref{eq:awa2} reads
  \begin{equation}
      \left(\overline{(aw_a)^2}\right)_{\ell m}\equiv\sum_iw_i^2\frac{a_{1,i}^2+a_{2,i}^2}{2}Y_{\ell m,i}^*.
  \end{equation}
  With this choice, the auto-spectrum of $\overline{(aw_a)^2}$ must be corrected by the following additive constant:
  \begin{equation}
    \tilde{N}_{\overline{(aw_a)^2}}=\frac{1}{4\pi}\sum_iw_i^4\left(\frac{a_{1,i}^2+a_{2,i}^2}{2}\right)^2.
  \end{equation}
  This generalises Eq. \ref{eq:4point_corr_spin0} for non-scalar fields.

\section{De-biasing \texorpdfstring{$\tilde{C}_\ell^{\overline{(aw_a)^2}}$}{the field-variance PCL}}\label{app:awa2_shotnoise}
  We derive here the bias to the pseudo-$C_\ell$ of the local variance map $\overline{(aw_a)^2}$, defined in Eq. \ref{eq:self_var}, when the local variance $\sigma_i^2\equiv \langle a_i^2\rangle$ is estimated directly from the data, and accounting for the subtraction of $\tilde{N}_a$ when estimating the auto-spectra of catalogue-based fields.

  \subsection{Correcting for \texorpdfstring{$\langle a_i^2\rangle\neq a_i^2$}{data-based point correlation estimates}}
    The local variance field $\overline{(aw_a)^2}$, entering the noise-noise component of the covariance matrix (Eq. \ref{eq:main_result}) through its pseudo-$C_\ell$ $\tilde{C}_\ell^{\overline{(aw_a)^2}}$, is defined in  Eq. \ref{eq:self_var} in terms of the point-wise variance $\langle a_i^2\rangle$. For a given catalogue, assuming we do not have an a priori estimate of $\langle a_i^2\rangle$, we simply replace it with its point estimate, $a_i^2$. The data-based local variance map is then given by Eq. \ref{eq:awa2}. The ensemble average of the pseudo-$C_\ell$ of this map (denoted here with a hat to stress that we used the data estimate $a_i^2\neq\langle a_i^2\rangle$) is:
    \begin{align}\nonumber
      \left\langle\hat{\tilde{C}}^{\overline{(aw_a)^2}}_\ell\right\rangle
      &= \sum_{ij}w_i^2w_j^2\langle a_i^2a_j^2\rangle \frac{1}{2\ell+1}\sum_mY^*_{\ell m,i}Y_{\ell m,j}\\\nonumber
      &= \sum_{ij}w_i^2w_j^2\langle a_i^2\rangle\langle a_j^2\rangle \frac{1}{2\ell+1}\sum_mY^*_{\ell m,i}Y_{\ell m,j}+2\sum_iw_i^4\langle a_i^2\rangle^2\,\frac{1}{2\ell+1}\sum_m|Y^*_{\ell m,i}|^2\\\label{eq:pcl_var_corr1}
      &=\tilde{C}^{\overline{(aw_a)^2}}_\ell+2\frac{1}{4\pi}\sum_iw_i^4\langle a_i^2\rangle^2 \, ,
    \end{align}
    where we have expanded the four-point cumulant in the second line using Wick's theorem, assuming uncorrelated field values ($\langle a_ia_j\rangle=\delta_{ij}\langle a_i^2\rangle$). The pseudo-$C_\ell$ of the data-based local variance map is thus biased by the second term in the equation above. We must therefore correct for this term, which we can do, again, through a data-based estimate of the field's squared variance $\langle a^2_i\rangle^2$. Since, for Gaussian fields, $\langle a_i^4\rangle=3\langle a_i^2\rangle^2$, this data-based correction can be calculated as
    \begin{equation}
      \Delta\tilde{C}^{\overline{(aw_a)^2},1}_\ell=\frac{2}{3}\frac{1}{4\pi}\sum_iw_i^4a_i^4 \, .
    \end{equation}

  \subsection{Correcting for \texorpdfstring{$\tilde{N}_a$}{field shot noise}}
    Our derivation of the Gaussian iNKA covariance in Section \ref{ssec:theo.general} assumed, for simplicity, that we are interested in the covariance of the pseudo-$C_\ell$ $\tilde{C}^{ab}_\ell$. While in general this is true, as we described in Section \ref{ssec:pcls.cat}, the auto-correlation of catalogue-based fields must be corrected for the contribution from self-pairs $\tilde{N}_a$ (Eq. \ref{eq:Na}). For auto-correlations, the quantity of interest is thus the combination
    \begin{equation}
      \tilde{C}^{aa,{\rm corr}}_\ell\equiv \tilde{C}^{aa}_\ell-\tilde{N}_a=\frac{1}{2\ell+1}\sum_m|\tilde{a}_{\ell m}|^2-\frac{1}{4\pi}\sum_i(w_ia_i)^2.
    \end{equation}
    The covariance of this quantity is then:
    \begin{align}\nonumber
      {\rm Cov}(\tilde{C}^{aa,{\rm corr}}_\ell,\tilde{C}^{aa,{\rm corr}}_{\ell'})=\frac{1}{(2\ell+1)(2\ell'+1)}&\left[\sum_{mm'}\langle |\tilde{a}_{\ell m}|^2|\tilde{a}_{\ell' m'}|^2\rangle-\frac{2\ell+1}{4\pi}\sum_{m'}\sum_iw_i^2\langle a_i^2|\tilde{a}_{\ell' m'}|^2\rangle\right.\\\nonumber
      &\left.-\frac{2\ell'+1}{4\pi}\sum_m\sum_iw_i^2\langle a_i^2|\tilde{a}_{\ell m}|^2\rangle+\frac{(2\ell+1)(2\ell'+1)}{(4\pi)^2}\sum_{ij}w_i^2w_j^2\langle a_i^2a_j^2\rangle
      \right].
    \end{align}
    Using Wick's theorem on all the 4-point correlators above for a Gaussian, noise-like field ($\langle a_ia_j\rangle=\delta_{ij}\langle a_i^2\rangle$), the expression above can be simplified, after a bit of algebra, into:
    \begin{align}
      {\rm Cov}(\tilde{C}^{aa,{\rm corr}}_\ell,\tilde{C}^{aa,{\rm corr}}_{\ell'})
      &=2\left[\Xi_{\ell\ell'}\left(\overline{(aw_a)^2},\overline{(aw_a)^2}\right)-\frac{1}{(4\pi)^2}\sum_i w_i^4\langle a_i^2\rangle^2\right].
    \end{align}
    This is equivalent to the noise-noise contribution in Eq. \ref{eq:main_result}, corrected by the second term above. Further insight into this correction may be gained by expanding the coupling coefficient $\Xi_{\ell\ell'}$ in terms of the pseudo-$C_\ell$ of its arguments (see Eq. \ref{eq:Xi}):
    \begin{align}\nonumber
      {\rm Cov}(\tilde{C}^{aa,{\rm corr}}_\ell,\tilde{C}^{aa,{\rm corr}}_{\ell'})
      &=2\left[\sum_{\ell''}\frac{2\ell''+1}{4\pi}\wtj{\ell}{\ell'}{\ell''}{0}{0}{0}^2\,\tilde{C}^{\overline{(aw_a)^2}}_{\ell''}-\frac{1}{(4\pi)^2}\sum_i w_i^4\langle a_i^2\rangle^2\right]\\
      &=2\sum_{\ell''}\frac{2\ell''+1}{4\pi}\wtj{\ell}{\ell'}{\ell''}{0}{0}{0}^2\left[\tilde{C}^{\overline{(aw_a)^2}}_{\ell''}-\frac{1}{4\pi}\sum_i w_i^4\langle a_i^2\rangle^2\right],
    \end{align}
    where, in the second line, we have used the property
    \begin{equation}
      \sum_{\ell''}(2\ell''+1)\wtj{\ell}{\ell'}{\ell''}{0}{0}{0}^2=1.
    \end{equation}

    We thus see that, to account for the $\tilde{N}_a$ subtraction in auto-spectra, the power spectrum of the local variance field entering the noise-noise covariance must be corrected by a term that is half of the correction derived in the previous section, which accounts for the fact that the local variance map is estimated from the data. Note that this correction is exactly the contribution of $\overline{(aw_a)^2}$ to the pseudo-$C_\ell$ from self-pairs. That is, it plays the same role as $\tilde{N}_a$ and $\tilde{N}_w$, accounting for the ``shot noise'' contribution to their corresponding power spectra. As in the previous section, we can estimate this term from the data assuming the field is Gaussian, obtaining the second correction
    \begin{equation}
      \Delta\tilde{C}^{\overline{(aw_a)^2},2}_\ell=\frac{1}{3}\frac{1}{4\pi}\sum_iw_i^4a_i^4.
    \end{equation}

    Thus, we find that when calculating the noise-noise contribution to the covariance matrix of auto-correlations, the pseudo-$C_\ell$ of the local variance map must be corrected by subtracting the quantity
    \begin{equation}
      \Delta\tilde{C}^{\overline{(aw_a)^2},1}_\ell+\Delta\tilde{C}^{\overline{(aw_a)^2},2}_\ell=\frac{1}{4\pi}\sum_iw_i^4a_i^4\equiv\tilde{N}_{\overline{(aw_a)^2}},
    \end{equation}
    defined in Eq. \ref{eq:4point_corr_spin0}.

\section{Numerical implementation of the brute force calculation} \label{app:brute_code}
In this section, we explain the code implementation and optimisation to compute the covariance matrix with exact summation. We implement the calculation using the \texttt{jax} library and design the code to run optimally on a GPU. The main advantage is the ability to leverage multiple \texttt{jax} properties to accelerate covariance computation without approximation.

The bottleneck of this method is calculating the high-dimensional array of Legendre polynomials at the pairs of source positions $P_{\ell, ij}$, which has dimensions $\ell_{max} \times N_{\text{source}}^2$, in addition to the expensive calculation of the Legendre polynomials for large $\ell$. For the former problem, we leverage the \texttt{jax} package to accelerate matrix computations. For the latter, we use the recursion relation of the Legendre Polynomials, known as the Bonnet recurrence formula  to produce recursive calculations of the Legendre polynomial \citep{Hademenos_Spiegel_Liu_2001}
\begin{equation}
    P_{n}(x) = \frac{2n-1}{n} \, x P_{n-1}(x) - \frac{n - 1}{n} P_{n-2}(x).
\end{equation}
For consecutive calculation, we use \texttt{jax.lax} property to loop over the calculation for $\ell_\mathrm{min} \leq \ell \leq \ell_\mathrm{max}$. In addition, to compute the correlator brackets given in Eq. \ref{eq:cov_brute}, we bin the estimated spectra into sets of multipoles (known as bandpowers) $b$ by assuming that the value of the power spectrum is constant for different multipoles for a particular bandpower, i.e.,
\begin{equation}
    \sum_{\ell} \frac{2\ell+1}{4\pi} C_\ell P_{\ell,ij}  \approx \sum_{b} C_b \sum_{\ell \in b} \frac{2\ell+1}{4\pi}  P_{\ell,ij}.
\end{equation}
When calculating a binned covariance matrix, we introduce a binning matrix based on the given multipole edges and apply it to the Legendre polynomials $P_{\ell, ij}$ before summing over the sources, thereby reducing the number of operations. In addition, for consistency with the numerical implementation of {\texttt{NmtWorkspace.decouple\_cell}}, we first bin the mode coupling matrix, then unbin it by assigning a constant value to each bandpower, and finally invert it in bandpower space, namely 
\begin{equation}
    [\mathcal{M}^{-1}]_{b\ell} = \sum_{b'}(B M U)^{-1}_{bb'} B^\ell_{b'}
\end{equation}
with $B$ being the binning matrix, $\sum_{\ell\in b}B_b^\ell=1$, and $U$ is the ``unbinning'' matrix, $\sum_b U^\ell_b C_b \equiv \sum_b C_b \Theta(\ell\in b)$, where $\Theta$ is a binary step function. The result is
\begin{equation}
    \mathrm{Cov}(\hat{C}_b^{aa}, \hat{C}_{b'}^{aa})  =  2
    \sum_{i \not= j,\, p \not= q}\frac{w_i w_jw_p w_q}{(4\pi)^2}   [\tilde{M}^{-1}P]_{b, ij} [\tilde{M}^{-1} P]_{b', pq}
     \left\langle a_i a_p \right\rangle \left\langle a_j a_q \right\rangle.
\end{equation}

Depending on the available GPU memory and the number of sources, it is possible to accommodate different memory constraints by splitting the sum into smaller parts. The code for performing exact summation for covariance matrix calculations is available in  the {\tt cpcl\_gpu}\footnote{\url{https://github.com/elyas-farah/cpcl_gpu.git}} repository. The repository contains two implementations of the covariance matrix: {\tt batch\_covariace} which parallelizes the exact summation with {\tt jax.lax} functionality, and another implementation, {\tt sum\_matrices} which uses dynamic for-loops to split the sum into smaller parts, surpassing memory restrictions on the expense of longer computational time.

\section{Reduction of the direct summation covariance to the NKA limit}\label{app:brute_nka}
The brute-force expression, Eq.~\ref{eq:cov_brute}, is exact for Gaussian fields, whereas the analytical covariance of Sect.~\ref{sec:theo} relies on the NKA. Here we show that the former reduces to the latter in the appropriate limit.
Using the addition theorem, the direct sum can be expressed through the masked harmonic coefficients $\tilde{a}_{\ell m}$ as:
\begin{equation}\label{eq:brute_as_cov}
    \mathrm{Cov}(\hat{C}_\ell^{aa},\hat{C}_{\ell'}^{aa}) = \sum_{\ell_1,\ell_2}[M^{-1}]_{\ell\ell_1}[M^{-1}]_{\ell_2\ell'}\,\mathrm{Cov}(\tilde{C}_{\ell_1}^{aa},\tilde{C}_{\ell_2}^{aa}),
\end{equation}
where
\begin{equation}\label{eq:pcl_cov_harm}
    \mathrm{Cov}(\tilde{C}_{\ell_1}^{aa},\tilde{C}_{\ell_2}^{aa}) = \frac{2}{(2\ell_1+1)(2\ell_2+1)}\sum_{m_1 m_2}\left|\langle \tilde{a}_{\ell_1 m_1}\tilde{a}^*_{\ell_2 m_2}\rangle\right|^2.
\end{equation}
This is the Gaussian pseudo-$C_\ell$ covariance of Sect.~\ref{ssec:pcls.cov}. The brute-force result is therefore the decoupled pseudo-$C_\ell$ covariance evaluated without approximation. The restriction to distinct source pairs ($i\neq j$, $p\neq q$) corresponds to the self-pair debiasing already applied in Eq.~\ref{eq:invert}.
 
Inserting the field two-point function, Eq.~\ref{eq:field_corr}, into the spherical harmonic expansion of the masked field separates the distinct-source signal from the self-pair shot noise. As in Sect.~\ref{sec:theo}, the distinct ($i\neq j$) pairs carry only the signal and, under the NKA, produce the self-pair-subtracted mask $\bar{w}_a^2-\overline{w_a^2}$ together with the iNKA spectrum $\overline{C}^{aa}$ of Eq.~\ref{eq:clbar0}, while the self-pairs ($i=j$) carry the full local variance $\langle a_i^2\rangle$ and assemble into the shot-noise map $\overline{(aw_a)^2}$ of Eq.~\ref{eq:awa2},
\begin{equation}\label{eq:masked_corr}
    \langle\tilde{a}_{\ell_1 m_1}\tilde{a}^*_{\ell_2 m_2}\rangle \approx \overline{C}^{aa}_{(\ell_1,\ell_2)}\,{\cal M}_{\ell_1 m_1,\ell_2 m_2}\!\left(\bar{w}_a^2-\overline{w_a^2}\right) + {\cal M}_{\ell_1 m_1,\ell_2 m_2}\!\left(\overline{(aw_a)^2}\right).
\end{equation}
Here we have pulled the slowly-varying $C^{aa}_{\ell''}\to\overline{C}^{aa}_{(\ell_1,\ell_2)}$ out of the convolution. Squaring Eq.~\ref{eq:masked_corr} and using
\begin{equation}\label{eq:M2_xi}
    \sum_{m_1 m_2}{\cal M}_{\ell_1 m_1,\ell_2 m_2}(u)\,{\cal M}^*_{\ell_1 m_1,\ell_2 m_2}(v) = (2\ell_1+1)(2\ell_2+1)\,\Xi_{\ell_1\ell_2}(u,v),
\end{equation}
Eq.~\ref{eq:pcl_cov_harm} becomes
\begin{equation}\label{eq:brute_nka}
\begin{split}
    \mathrm{Cov}(\tilde{C}_{\ell}^{aa},\tilde{C}_{\ell'}^{aa}) \approx 2\Big[&\big(\overline{C}^{aa}_{(\ell,\ell')}\big)^2\,\Xi_{\ell\ell'}\!\left(\bar{w}_a^2-\overline{w_a^2},\,\bar{w}_a^2-\overline{w_a^2}\right) + 2\,\overline{C}^{aa}_{(\ell,\ell')}\,\Xi_{\ell\ell'}\!\left(\bar{w}_a^2-\overline{w_a^2},\,\overline{(aw_a)^2}\right) \\
    &+ \Xi_{\ell\ell'}\!\left(\overline{(aw_a)^2},\,\overline{(aw_a)^2}\right)\Big],
\end{split}
\end{equation}
which is Eq.~\ref{eq:main_result} for $a=b=c=d$.

\twocolumngrid

\bibliography{main}

\end{document}